\documentclass[reprint,prd,onecolumn,notitlepage,11pt,nofootinbib]{revtex4-1}
\usepackage[english]{babel}
\usepackage{amsmath,amssymb,graphicx,bm,lipsum}
\usepackage[colorlinks=true,citecolor=blue,linkcolor=blue,urlcolor=blue]{hyperref} 
\usepackage[font={small},flushleft,indent]{caption}
\usepackage{setspace}

\begin{document}

\title{Observational signatures from higher-order images of moving hotspots in accretion disks}

\author{Qing-Hua Zhu}
\email{zhuqh@cqu.edu.cn} 
\affiliation{School of Physics, Chongqing University, Chongqing 401331, China} 
 
\begin{abstract}  
The efforts to probe the horizon-scale structure of black holes, such as Event Horizon Telescope and GRAVITY interferometer, might provide valuable insights into the strong-field regime of Einstein's theory of gravity. In the near field region of a black hole, the observational signatures of moving hotspots might potentially reveal the mechanism causing the flares, or reflect the spacetime geometries. This paper develops a ray tracing scenario to study higher-order images of moving hotspots in a thin disk around a spherical black hole. Our ray-tracing scenario establishes a one-to-one mapping between emission locations and observer's sky. It enables us to perform infinite-precision simulations for the images, because the emission sources are projected directly onto the image plane. Furthermore, we show that a source located anywhere outside the black hole can be repeatedly mapped onto the observer's sky, from primary to higher-order images. We investigate the observational signatures of hotspots, focusing on temporal fluxes and flux centroids from the primary to sixth-order images. The hotspots are considered to be moving in circular, escape, and plunging orbits. Our results find that the higher-order images can be categorized into two types. Within each type, the temporal fluxes exhibit a self-similar profile. Furthermore, as the hotspots approach the event horizon of a black hole, the fluxes from higher-order images alternately dominate the observed flux, which subsequently result in the flux decaying with time in an oscillatory manner.

\end{abstract} 

\maketitle

\section{Introduction}
The Event Horizon Telescope (EHT) captured the first image of the supermassive black hole in M87 \cite{EventHorizonTelescope:2019ths,EventHorizonTelescope:2019dse} and subsequently Sgr A* \cite{EventHorizonTelescope:2022wkp,EventHorizonTelescope:2022wok}, both of which showed a ring-like appearance. On the other side, the GRAVITY interferometer can measure the position of the center of light with high precision and reported flare events near Sgr A* \cite{GRAVITY:2018sef,GRAVITY:2023avo}. These efforts to probe the horizon-scale structure of black holes might provide valuable insights into the strong-field regime of Einstein's theory of gravity \cite{Narayan:2023ivq}.
 
The flare events are reported within five Schwarzschild radii from the supermassive black hole at the galactic center \cite{GRAVITY:2018sef,GRAVITY:2023avo}, and thus might reflect properties of accretion matter in the strong-field regime of gravity. These flares are expected to originate from magnetohydrodynamic effects, such as magnetic reconnection \cite{Dexter:2020cuv,Aimar:2023kzj}, or dynamical processes of accretion flow \cite{Baganoff:2001kw,Genzel:2003as,Yuan:2003dc,GRAVITY:2021hxs}. The observations in X-ray and near-infrared wavelengths indicated that the flare radiation flux is highly polarized and quasi-periodic, last for about an hour \cite{Genzel:2003as,2006A&A...458L..25M,2006A&A...460...15M,Zamaninasab:2009df,Pechacek:2008yf}. Recently, GRAVITY collaboration reported that position centroid of the flare exhibits clockwise orbital motion on the sky \cite{GRAVITY:2018sef,GRAVITY:2023avo}. It can be modeled as a 'hotspot' moving around the black hole \cite{GRAVITY:2020hwn,GRAVITY:2020lpa}. And this model was characterized by predicting a shift in the image centroid \cite{Broderick:2005at,Zamaninasab:2009df}. While point-like objects, like the star S2, are expected to undergo geodesic motion near the galactic center \cite{GRAVITY:2020gka}, a single hotspot might not move in geodesic motion, as indicated by the observations \cite{GRAVITY:2018sef,GRAVITY:2020lpa,Matsumoto:2020wul,GRAVITY:2023avo,Antonopoulou:2024qco}. The deviation from geodesic motion, as revealed by the hotspot behavior, might offer insights into the mechanisms causing the flares.

In this context, the hotspot is a phenomenological model describing the emission region near a black hole. The hotspots were assumed to be distributed on the surface of accretion disks to explain the X-ray variability observed in active galactic nuclei \cite{1992A257594B,1991A245454A,1990ComAp,Garcia:2015nmh}. Later, this model was extended to infrared wavelengths \cite{Broderick:2005jj,Broderick:2005my,Do:2008uf,GRAVITY:2017fis}. In these studies, the observational signatures such as temporal fluxes and flux centroids \cite{Zamaninasab:2009df,Li:2014coa}, obtained using ray-tracing algorithm, were investigated. Recently, by considering effect from black hole spin \cite{Bozza:2010xqn,Martocchia:1999sn,1994MNRAS.269..283Z}, various spacetime geometries \cite{Macedo:2024qky,Tamm:2023wvn,Chen:2024ilc,Chen:2023knf,Rosa:2023qcv,Rosa:2022toh,Liu:2014awa,Li:2014fza,Li:2014coa,Zeng:2024ptv} and different orbits of the hotspots \cite{Vincent:2014nja,GRAVITY:2023avo,Antonopoulou:2024qco,Huang:2024wpj}, distinctive observational signatures of the hotspots were also identified.

For the simulations of black hole imaging, the conventional ray tracing algorithm works by tracing the path of light from each pixel on the image plane until it reaches the emission sources \cite{1992MNRAS.259..569K,Dexter:2009fg,Vincent:2011wz,Dexter:2016cdk}. It is effective, if the emission sources occupy major portion of the image plane. However, it might have a high computational cost for small or even point-like sources, because most light rays never reach the sources.  In order to avoid this situation, one might expect to determine the light paths between the locations of emission sources and observers. In fact, this scenario was implemented decades ago to study the temporal fluxes from the lower-order images of a corotating hotspot in accretion disks  \cite{1992A257594B,1994ApJ...425...63B}, including the influence of eclipses \cite{1992A&A...257..531K}. Recently, it was employed in so-called adaptive ray tracing to determine the lensing bands \cite{Paugnat:2022qzy, Cardenas-Avendano:2022csp}. An alternative formalism was also presented in Ref.~\cite{Zhu:2023kei}. It was also extended for studying astrometry of the higher-order image of static hotspots, and is named ``forward ray tracing'' \cite{Zhou:2024dbc}.

This paper extends the ray tracing scenario proposed in Refs.~\cite{1992A257594B,1994ApJ...425...63B}, where our formulae can be applicable to higher-order images and  for the emission sources beyond equatorial plane. We study moving hotspots in a thin disk around spherical black holes, and focus on their observational signatures such as temporal fluxes and flux centroids, rather than on the astrometry \cite{Wielgus:2021peu,Bisnovatyi-Kogan:2022ujt,Tsupko:2022kwi,Wang:2022mjo,Kocherlakota:2024hyq,Zhou:2024dbc}. With fixed observer locations, our ray-tracing scenario establishes a one-to-one mapping between the emission locations and the observer's sky. It enables us to perform infinite-precision simulations for the images, because the emission sources are projected directly onto the image plane. Furthermore, we show that a source located anywhere outside the black hole can be repeatedly mapped onto the observer's sky, from  primary to higher-order images. We present the observational signatures of the hotspots from primary to sixth-order images. These hotspots are considered moving in circular, escape, and plunging orbits for given initial radii and initial azimuth angles. Our results find that the higher-order images can be categorized into two types. Within each type, the temporal fluxes exhibit a self-similar profile. For illustration, we consider the Schwarzschild de-Sitter black hole. It shows that the profiles of temporal fluxes can be influenced by the orbital angular velocities and radii but are little affected by the cosmological constant. And the centroid positions are shown to be distinctive from the tracks of the primary images, especially when the hotspots pass behind the black hole.

The rest of the paper is organized as follows. In Sec.~\ref{II}, we introduce  observers' sky and images plane based on astrometric approach \cite{Chang:2020miq,Zhu:2023kei}. In Sec.~\ref{III}, we present the scenario of ray tracing that create a one-to-one mapping between locations of the emission sources and the observers' sky. In Sec.~\ref{IV}, we study observational signatures of hotspots in a thin disk, from primary to sixth-order images, considering the hotspots in circular, escape, and plunging orbits. In Sec.~\ref{V}, conclusions and discussions are summarized.

\ 

\section{Observers' sky and Image plane for spherical black holes \label{II}}

As a preliminary, we list the known results of light propagation formulated by geodesic equations, and introduce the image plane following the astrometric approach developed in Refs.~\cite{Chang:2020miq,Zhu:2023kei}. 
We consider a spherical black hole, where the metric can be given by
\begin{eqnarray}
  \textrm{d} s^2 & = & - f (r) \textrm{d} t^2 + \frac{\textrm{d} r^2}{f (r)} + r^2 (\textrm{d}  \theta^2 + \sin\theta^2\textrm{d}\phi^2) ~. \label{met}
\end{eqnarray}
For the static and spherical spacetime, the
4-momentum of light ray $k_{\mu}$ takes the form of
\begin{eqnarray}
    k_t  =  - E ~, \hspace{0.5cm}
    k_r  =  \pm_r \frac{1}{f} \sqrt{E^2 - \frac{K   f  }{r^2}}~, \hspace{0.5cm}
    k_{\theta}  =  \pm_{\theta} \sqrt{K - \frac{L^2}{\sin^2 \theta}} ~, \hspace{0.5cm}
    k_{\phi}  =  L ~, \label{mom}
\end{eqnarray}
where  $K$ is Carter's constant, the $L$ represents the angular
momentum with respect to axis-$z$ and $E$ is the energy of the light. The sign functions $\pm_r$ and $\pm_\theta$ change from $\pm$ to $\mp$ when a light ray encounters a turning point in radius and inclination, respectively. 
As the $k_{\theta}^2 > 0$ leads to $K \geq L^2 / \sin^2 \theta \geq0$, we introduce the quantities $\rho$
and $\varphi$ as follows,
\begin{eqnarray}
  \rho  \equiv  \frac{\sqrt{K}}{E}~, &&
  \cos \varphi  \equiv  - \frac{L}{\sqrt{K} \sin \theta_\text{o}} ~, \label{rhovarphi}
\end{eqnarray}
where $x_\text{o} [\equiv (t_\text{o}, r_\text{o}, \theta_\text{o}, \phi_\text{o})]$ denotes as the event of observer.

Employing $\rho$ and $\varphi$ in Eq.~(\ref{rhovarphi}), we evaluate the 4-momentum of light ray in Eqs.~(\ref{mom}) as a set of integrals, namely,
\begin{subequations}
  \begin{eqnarray}
    0 & = & \mp_{\theta} \int_{\mathcal{C}} \frac{\textrm{d} \theta}{\sqrt{1 -
    \csc^2 \theta \sin^2 \theta_\text{o} \cos^2 \varphi}} \pm_r \rho \int_{\mathcal{C}}
    \textrm{d} r \left\{ \frac{1}{r \sqrt{r^2 - \rho^2 f}} \right\}~, \label{intgeo1} \\
    t_\text{o} - t_\text{s} & = & \pm_r \int_{\mathcal{C}} \textrm{d} r \left\{ \frac{r}{f
    \sqrt{r^2 - \rho^2 f}} \right\} ~,\label{intgeo2}\\
    \phi_\text{o} - \phi_\text{s} & = & \mp_{\theta} \sin \theta_\text{o} \cos \varphi
    \int_{\mathcal{C}} \textrm{d} \theta \left\{ \frac{1}{\sin^2 \theta \sqrt{1 -
    \csc^2 \theta \sin^2 \theta_\text{o} \cos^2 \varphi}} \right\} ~,\label{intgeo3}
  \end{eqnarray} \label{intgeo}
\end{subequations}
where $x_\text{s} [\equiv (t_\text{s}, r_\text{s}, \theta_\text{s}, \phi_\text{s})]$ represents the event of
the emission source, and the integral is performed along the worldline of light
$\mathcal{C}$. From Eq.~(\ref{intgeo}), one might find that the $\rho$ and $\varphi$ are separated in the integrals of $r$ and $\theta$.

The  observers' sky can be established with local frame adapted to zero angular momentum observers (ZAMOs) \cite{Bardeen:1973tla}.  However, both the local frame and finite-distant ZAMOs might not be physical objectives.  In observations, the concept of astrometry is developed by astronomers positing the star on the sky, according to known reference objects in the universe. For examples, International Celestial Reference System is based on hundreds of extra-galactic sources distributed on the sky \cite{Gaia:2018}. And the
astrometry of flare events near the Sgr A* are located with the relative position of star S2 \cite{GRAVITY:2018sef}. The observer's sky seems inevitably to be defined with the physical objectives.
In a theoretical context, this scenario has been utilized in the field of black hole images, so-called astrometric approach \cite{Chang:2019vni,Chang:2020miq,Chang:2020lmg,He:2020dfo,Chang:2021ngy,Guo:2022nto,Zhu:2023kei}. Here, we further provide a simpler and improved formalism for spherical spacetime.

We adopt two reference light rays from distant stars $k^{(\theta)}$, $k^{(\phi)}$, and a reference light from the black hole center $k^{(r)}$, namely,
\begin{subequations}
  \begin{eqnarray}
    k^{(r)} & = & E^{(r)} \left( - \textrm{d} t + \frac{1}{f} \textrm{d} r \right)~, \\
    k^{(\theta)} & = & E^{(\theta)} \left( - \textrm{d} t + \frac{1}{f} \sqrt{1 -
    \left( \frac{r_\text{o}}{r} \right)^2 \frac{f (r)}{f (r_\text{o})}} \textrm{d} r - \frac{r
     _\text{o}}{\sqrt{f (r_\text{o})}} \textrm{d} \theta \right) ~,\\
    k^{(\phi)} & = & E^{(\phi)} \left( - \textrm{d} t + \frac{1}{f} \sqrt{1 - \left(
    \frac{r_\text{o}}{r} \right)^2 \frac{f (r)}{f (r_\text{o})}} \textrm{d} r + \frac{r_\text{o}}{\sqrt{f
    (r_\text{o})}} \sqrt{1 - \frac{\sin^2 \theta_\text{o}}{\sin^2 \theta}} \textrm{d} \theta +
    \frac{r_\text{o} \sin \theta_\text{o}}{\sqrt{f (r_\text{o})}} \textrm{d} \phi \right) ~,
  \end{eqnarray} \label{reference}
\end{subequations}
where $E^{(r)}$, $E^{(\theta)}$ and $E^{(\phi)}$ are three independent integral constants (energy) of the 4-momentums $k^{(r)}$, $k^{(\theta)}$ and $k^{(\phi)}$, respectively. We will show later that the values of them have no influence on astrometry. 
Here, one might find that $k^{(r)}$, $k^{(\theta)}$ and $k^{(\phi)}$ are orthogonal to each other at $x_\text{o}$. In order to locate a source with its emitted light $k_{\mu}$, the astrometric observable can be obtained by introducing the spatial inner product as follows \cite{Chang:2019vni},   
\begin{eqnarray}
  \langle k_1, k_2 \rangle & \equiv & \frac{\gamma^{\ast} k_1}{|
  \gamma^{\ast} k_1 |} \cdot \frac{\gamma^{\ast} k_2}{|
  \gamma^{\ast} k_2 |} = 1 + \frac{k_1 \cdot k_2}{(u \cdot k_1) (u \cdot
  k_2)}~, \label{defAngle}
\end{eqnarray}
where $\gamma^{\ast}$ is the induced metric adapted to the coordinate time and `$\cdot$' denotes the metric contraction with respect to $g_{\mu \nu}$  in Eq.~(\ref{met}). By making use of Eq.~(\ref{defAngle}), the location of light ray $k_{\mu}$ on the sky can be expressed as celestial coordinate $(\Phi, \Psi)$,
namely,
\begin{subequations}
  \begin{eqnarray}
    \cos \Phi \sin \Psi & \equiv & \langle - k, k^{(\phi)} \rangle = \cos
    \varphi \frac{\rho \sqrt{f (r_\text{o})}}{r_\text{o}}~, \\
    \sin \Phi \sin \Psi & \equiv & \langle - k, k^{(\theta)} \rangle   =
    \sigma_{\theta} | \sin \varphi | \frac{\rho \sqrt{f (r_\text{o})}}{r_\text{o}}~, \\
    \cos \Psi & \equiv & \langle - k, k^{(r)} \rangle = - \sigma_r \sqrt{1 -
    \frac{\rho^2 f (r_\text{o})}{r_\text{o}^2}}~, 
  \end{eqnarray}\label{celestial1}
\end{subequations}
where $\sigma_{\theta}[ \equiv \text{sign} (k_{\theta} |_{x_\text{o}})]$ and $\sigma_r[\equiv \text{sign} (k_r |_{x_\text{o}})$] are the sign functions of $k_\theta$ and $k_r$ at $x_\text{o}$. By associating the conventions
$\sigma_{\theta} = \text{sign} (\varphi)$ and $\sigma_r = \text{sign} (\cos
\Psi)$ with Eqs.~(\ref{celestial1}), the celestial coordinate of $k_{\mu}$ can be rewritten as $\varphi$ and $\rho$, namely,
\begin{subequations}
  \begin{eqnarray}
    \Phi & = & \arctan \left( \frac{\langle k, k^{(\theta)} \rangle}{\langle k,
    k^{(\varphi)} \rangle} \right) = \varphi ~,\\
    \Psi & = & \arccos \langle k, k^{(r)} \rangle = \arccos \left( - \sigma_r
    \sqrt{1 - \frac{\rho^2 f (r_\text{o})}{r_\text{o}^2}} \right) ~.
  \end{eqnarray} \label{Image}
\end{subequations}
One can obtain $\sin \Psi = \rho \sqrt{f (r_0)} / r_\text{o}$, where the $\Psi$
 describes the astrometry related to the black hole center.

\begin{figure}
  \includegraphics[width=0.6\linewidth]{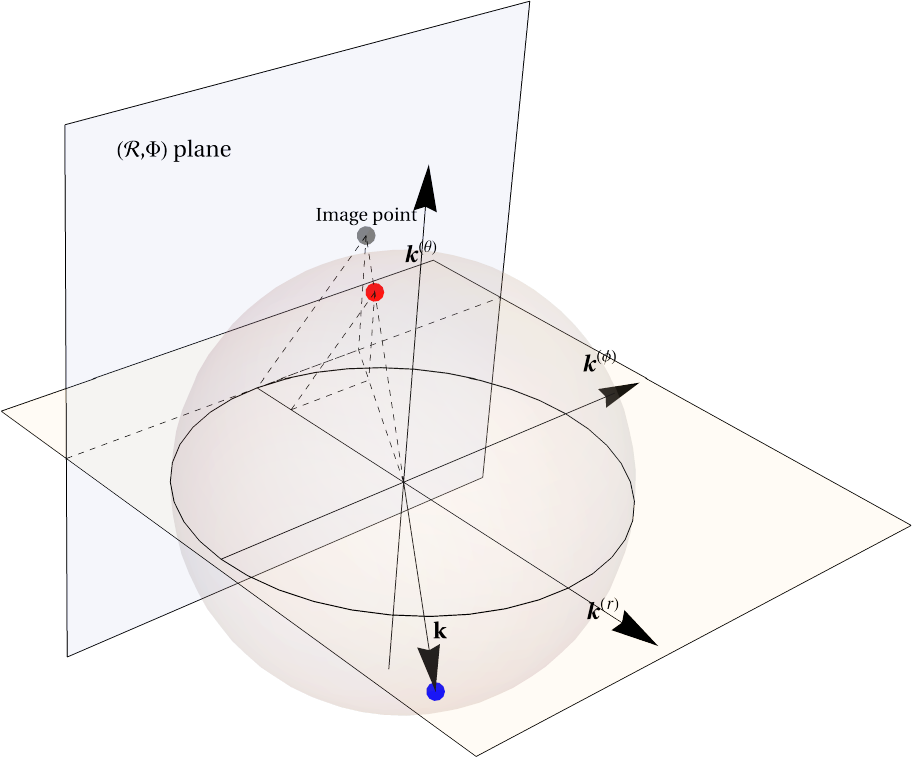} 
  \caption{Schematic diagram for illustrating observers' sky and image plane
   $(\mathcal{R},\Phi)$. The red and gray points denote the image points on celestial sphere and image plane, respectively. \label{F1}}
\end{figure}

For distant observers, one can utilize the projection plane composed of $(\mathcal{R},
\Phi)$, where 
\begin{eqnarray}
  \mathcal{R} & \equiv & \tan \Psi = \rho \sqrt{\frac{f (r_\text{o})}{r_\text{o}^2 - \rho^2 f
  (r_\text{o})}} ~. \label{ImagePol}
\end{eqnarray} 
We present schematic diagram about the observers' sky and the image plane in Figure~\ref{F1}.
In previous studies~\cite{Chang:2020miq, Chang:2020lmg, Chang:2021ngy, Zhu:2023kei}, the reference light rays were set to be those from the photon sphere. It follows the scenario where the observers' sky is established based on physical objectives, despite increasing the complexity in calculations.
 
\ 

\section{Determination of light rays for given ending and starting points \label{III}}

The conventional ray tracing scenario creates a mapping from observers'
sky to the emission source with given observers' location. Namely, path of light from each pixel on the image plane is traced backward until it reaches the
emission sources \cite{1992MNRAS.259..569K,Dexter:2009fg,Vincent:2011wz,Dexter:2016cdk}. It would be effective if the emission covers major portion of the image plane. However, when the emission sources are small or even point-like, the scenario has a high computational cost, as most light rays never reach the sources.
Besides the conventional ray tracing scenario $\text{CRT} : (x_\text{o}, \rho, \varphi) \mapsto x_\text{s}$, one can alternatively establish a mapping $\text{RT} : (x_\text{o}, x_\text{s}) \mapsto (\rho, \varphi)$. This scenario was implemented in Ref.~\cite{1992A257594B} decades ago in the study of corotating hotspots for primary images. Here, we will extend it to higher-order images.  

For rotating black holes, this scenario might not be that efficient, because the $\rho$ and $\varphi$ are coupled in the geodesic equations \cite{Zhu:2023kei}. The computation resources are spent in the procedure of root-finding in the two-dimensional plane \cite{Zhou:2024dbc}. Fortunately, there is no such difficulty for spherical black holes, as the integrals involving $\rho$ and $\varphi$ are decoupled as shown in Eqs.~(\ref{intgeo}). 
We can establish a mapping $\text{RT} : (x_\text{o}, x_\text{s}) \mapsto (\rho, \varphi)$, which allows the higher-order images of moving hotspots to be precisely and efficiently determined.

\begin{figure}
  \includegraphics[width=0.7\linewidth]{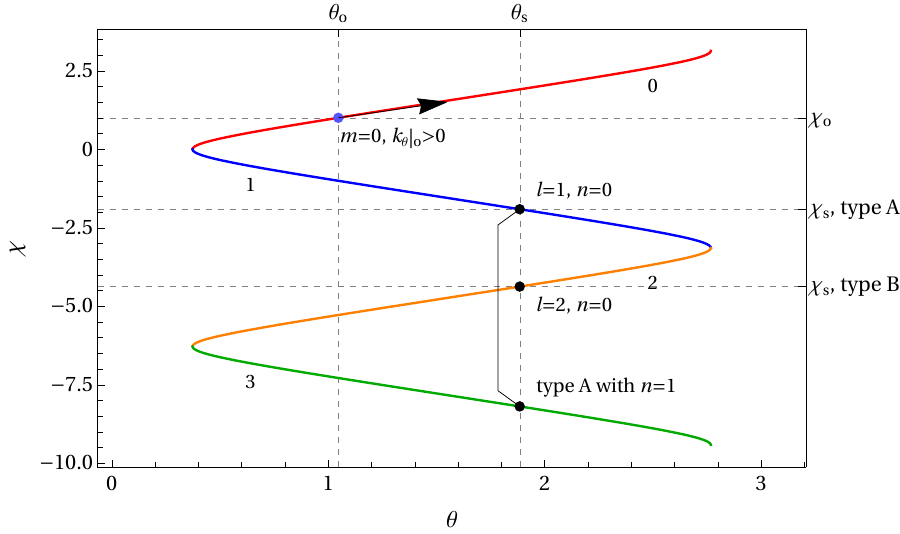}
  \caption{Relation between $\chi$ and $\theta$ for illustrating rule defined in Eq.~(\ref{rule}). The $\chi_\text{o}$ lies within the range of the red ($m=0$) or the blue ($m=1$) segments, while the $\chi_\text{s}$ with different $n$ and $l$ can take values in the range of segments in different colors.  \label{F2}}
\end{figure}

Firstly, we present the procedure of determination of the $\varphi$ with given
$\theta_\text{o}$ and $\theta_\text{s}$. Following the approach used in Ref.~\cite{Zhu:2023kei}, we adopt
the variable substitution, namely
\begin{eqnarray}
  \cos \theta & = & \sqrt{1 - \sin^2 \theta_\text{o} \cos^2 \varphi} \cos \chi ~. \label{chi}
\end{eqnarray}
Thus, the $\chi$ can be inversely given by
\begin{subequations}
  \begin{eqnarray}
    \chi_\text{o} & = & \left( \frac{1 - (- 1)^m}{2} - m \right) \pi + (- 1)^m \arccos
    \left( \frac{\cos \theta_\text{o}}{\sqrt{1 - \sin^2 \theta_\text{o} \cos^2 \varphi}}
    \right)~, \\
    \chi_\text{s} & = & - 2 \pi n + \left( \frac{1 - (- 1)^l}{2} - l \right) \pi + (-
    1)^l \arccos \left( \frac{\cos \theta_\text{s}}{\sqrt{1 - \sin^2 \theta_\text{o} \cos^2
    \varphi}} \right) ~,
  \end{eqnarray} \label{chiso}
\end{subequations}
where $m$ and $l$ are integers, the $n$ is the winding number of light orbits, the $\chi$ at location of source and observer are denoted as $\chi_\text{o}$ and
$\chi_\text{s}$, and we
adopt the convention $\chi_\text{o} \in [- \pi, \pi]$, and $\chi_\text{s} \in (- \infty,
\infty)$. By making use of the variable substitution in Eq.~(\ref{chi}), the first
term in Eq.~(\ref{intgeo1}) reduce to $\chi_\text{o} - \chi_\text{s}$. In order to ensure that the $\chi_\text{o} -
\chi_\text{s}$ remain positive, we establish the following rule for $m$ and $n$, namely,
\begin{equation}
  \left\{\begin{array}{ll}
    m = 0, l = 0 & (\varphi \geq 0) \cap (\theta_\text{o} \geq \theta_\text{s}) \cap  \textrm{A}
    \\
    m = 0, l = 1 &  ((\varphi \geq 0) \cap (\theta_\text{o} \geq \theta_\text{s}) \cap  \textrm{B}
    ) \cup ((\varphi \geq 0) \cap (\theta_\text{o} < \theta_\text{s}) \cap  \textrm{A} )\\
    m = 0, l = 2 & (\varphi \geq 0) \cap (\theta_\text{o} < \theta_\text{s}) \cap  \textrm{B} \\
    m = 1, l = 1 & (\varphi < 0) \cap (\theta_\text{o} < \theta_\text{s}) \cap  \textrm{A}
    \\
    m = 1, l = 2 & ((\varphi < 0) \cap (\theta_\text{o} < \theta_\text{s}) \cap  \textrm{B} )
    \cup ((\varphi < 0) \cap (\theta_\text{o} \geq \theta)_\text{s} \cap  \textrm{A} )\\
    m = 1, l = 3 & (\varphi < 0) \cap (\theta_\text{o} \geq \theta_\text{s}) \cap  \textrm{B} 
  \end{array}~.\right. \label{rule}  
\end{equation}
where we also adopted the convention, $\text{sign} (k_{\theta} |_{x_\text{o}}) =
 \text{sign} (\Phi)$, as introduced in Sec.~\ref{II}. Namely, when the light ray propagates as $\theta$ increases, it is imaged onto the northern hemisphere on the sky. The schematic diagram of Eq.~(\ref{rule}) is illustrated in Figure~\ref{F2}. 
 For given $\theta_\text{o}$ and $\theta_\text{s}$, one can not uniquely determine the $\chi_\text{s}$ with Eqs.~(\ref{chiso}) and (\ref{rule}), even in the case of $n=0$. Here, we introduce the A-type and B-type images, which are classified based on the closest and second-closest values of $\chi_\text{s}$ to the $\chi_\text{o}$. 

\begin{figure}
  \includegraphics[width=1\linewidth]{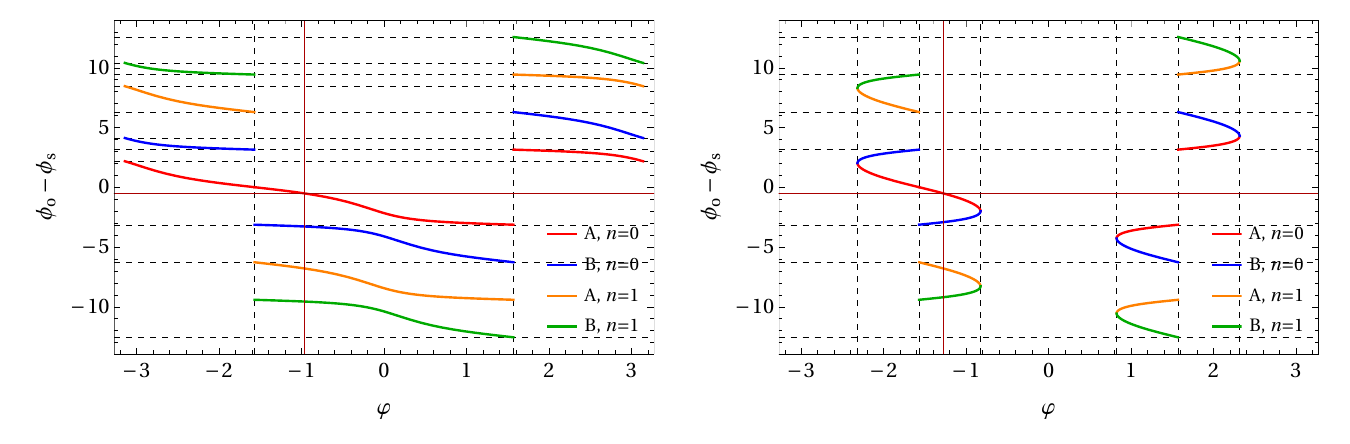}
  \caption{Relations between $\phi_\text{o} - \phi_\text{s}$ and $\varphi$ for illustrating Eq.~(\ref{phios}). We present the plots in the case of $\sin \theta_\text{s} \geq \sin \theta_\text{o}$ in left panel and $\sin \theta_\text{s} < \sin \theta_\text{o}$ in the right panel.  \label{F3}}
\end{figure}

The higher-order images of an emission source can be attributed to a non-zero $n$. It can be
derived from the values of $\phi_\text{s}$, if the range of it is
extended into $(- \infty, \infty)$. By making use of the substitution in
Eq.~(\ref{chi}), Eq.~(\ref{intgeo3}) can be evaluated to be
\begin{eqnarray}
  \phi_\text{o} - \phi_\text{s} 
  & = & - \arctan \left( \frac{\tan \varphi}{\cos \theta_\text{o}} \right) + (-
  1)^{l } \arctan \left( \frac{\sqrt{\sin^2 \theta_\text{s} - \sin^2 \theta_\text{o}
  \cos^2 \varphi}}{\sin \theta_\text{o} \cos \theta_\text{s} \cos \varphi} \right)
  \nonumber\\
  &  & - \pi \text{sign} ( \cos \varphi) \left( 2 n + l - m + \frac{(- 1)^l
  - (- 1)^m}{2} + (- 1)^m \left[ \frac{\theta_\text{o}}{\pi / 2} \right] - (- 1)^l
  \left[ \frac{\theta_\text{s}}{\pi / 2} \right] \right) ~, \label{phios}
\end{eqnarray}
where if $\sin \theta_\text{s} < \sin \theta_\text{o}$, we have $\varphi \in [- \varphi_-, -
\varphi_+] \cup [\varphi_+, \varphi_-]  $ and $\varphi_{\pm} \equiv
\arccos (\pm \sin \theta_\text{s} / \sin \theta_\text{o})$, otherwise, we have $\varphi \in
[- \pi, \pi]$. The square bracket $[ \hspace{0.2cm} ]$ in Eq.~(\ref{phios}) represents the floor function. 
In the case of $\sin \theta_\text{s} < \sin \theta_\text{o}$, it leads to
\begin{eqnarray} 
   (\phi_\text{o} - \phi_\text{s}) |_{\varphi = \varphi_{\pm}} & = & \pm \Bigg( (-
  1)^{m + 1} \arctan \left( \frac{1}{\cos \theta_\text{o}} \sqrt{\frac{\sin^2
  \theta_\text{o}}{\sin^2 \theta_\text{s}} - 1} \right) 
  \nonumber \\ && 
  - \pi \left( l - m + \frac{(- 1)^l - (- 1)^m}{2} + (- 1)^m \left[
  \frac{\theta_\text{o}}{\pi / 2} \right] - (- 1)^l \left[ \frac{\theta_\text{s}}{\pi / 2}
  \right] \right) \Bigg) ~,
\end{eqnarray}
In the case of $\sin \theta_\text{s} \geq \sin \theta_\text{o}$, we have
\begin{eqnarray}
  (\phi_\text{o} - \phi_\text{s})_{\varphi = \pm \pi} & = & (- 1)^{l + 1} \arctan \left(
  \frac{1}{\cos \theta_\text{s}} \sqrt{\frac{\sin^2 \theta_\text{s}}{\sin^2 \theta_\text{o}} - 1}
  \right) \nonumber\\
  &  & + \pi \left( 2 n + l - m + \frac{(- 1)^l - (- 1)^m}{2} + (- 1)^m
  \left[ \frac{\theta_\text{o}}{\pi / 2} \right] - (- 1)^l \left[ \frac{\theta_\text{s}}{\pi
  / 2} \right] \right) ~,
\end{eqnarray}
Here, we can adopt the convention, $\phi_\text{o} = 0$ for simplicity. In Figure~\ref{F3}, we
show  Eq.~(\ref{phios}) for illustrating that the $\varphi$ can be uniquely determined with given $\phi_{\rm o}-\phi_{\rm s}$. 

As illustration in Figures~\ref{F2} and \ref{F3}, we can determine A-type or B-type images, the winding number $n$, and the value of $\varphi$ with given $\theta_\text{o}$, $\theta_\text{o}$ and $\phi_\text{o} - \phi_\text{s}$. Namely, the scenario has created a mapping $(\theta_\text{s}, \theta_\text{o}, \phi_\text{o} - \phi_\text{s}) \mapsto \left(
\text{A-B type},   n, \varphi \right)$. It is a general result for
spherical black holes, as it is independent of the $f (r)$. In fact, the way to determine the $\varphi$ has been proposed to study the
primary image (image of type A with $n = 0$) of
corotating hotspots on equatorial plane ($\sin \theta_\text{s} \leq \sin \theta_\text{o} $ always holds) \cite{1992A257594B}.

We also present the procedure to determine $\rho$ with given $r_\text{o}$
and $r_\text{s}$. For illustration, we consider the Schwarzschild de-Sitter black
hole, where $f (r) = 1 - {2 M}/{r} - {\Lambda r^2}/{3}$. In this case, Eq.~(\ref{intgeo1}) can be evaluated to be
\begin{eqnarray}
  \rho (\mathcal{I} (r_\text{o} ; \rho) \pm \mathcal{I} (r_\text{s} ; \rho)) & = & \chi_\text{o} -
  \chi_\text{s} \nonumber\\
  & = & 2 \pi n + \left( \frac{(- 1)^l - (- 1)^m}{2} + l - m \right) \pi  + (- 1)^l \arccos \left( \frac{\cos \theta_\text{s}}{\sqrt{1 - \sin^2
  \theta_\text{o} \cos^2 \varphi}} \right) \nonumber\\
  &  & + (- 1)^m \arccos \left( \frac{\cos
  \theta_\text{o}}{\sqrt{1 - \sin^2 \theta_\text{o} \cos^2 \varphi}} \right)~, \label{ros}
\end{eqnarray}
where $l$ and $m$ can be obtained by making use of rule in Eq.~(\ref{rule}) with
known ($\theta_\text{o}$, $\theta_\text{s}$, $\text{A-B type}$, $\varphi$), and the
$\mathcal{I} (r ; \rho)$ is the result of analytical integral in the second term of Eq.~(\ref{intgeo1}), namely
\begin{eqnarray}
  \mathcal{I} (r ; \rho) & \equiv & \left\{\begin{array}{ll}
    \frac{2}{\Xi \sqrt{(r_3 - r_1) r_2}} F \left( \arcsin \sqrt{\frac{r_2 (r -
    r_3)}{r_3 (r - r_2)}} \Big| \frac{r_3 (r_2 - r_1)}{r_2 (r_3 - r_1)} \right)
    ~, & \rho \geq \rho_c\\
    \frac{2}{\Xi \sqrt{(r_3 - r_1) r_2}} F \left( \arcsin \sqrt{\frac{r (r_3 -
    r_1)}{r_3 (r - r_1)}}\Big| \frac{r_3 (r_2 - r_1)}{r_2 (r_3 - r_1)} \right)
    ~, & \rho < \rho_c
  \end{array}\right. ~. \label{Ir}
\end{eqnarray}
The $F(x|y)$ is the elliptic integral of the first kind, and we set $\rho_c \equiv 3 \sqrt{3} M / (1 - 9 \Lambda
M^2)$, $\Xi \equiv \sqrt{1 + \Lambda \rho^2 / 3}$. The $r_1$, $r_2$ and
$r_3$ are three roots of equation $r^2 - \rho^2 f = 0$. 
The $\pm$ in lhs. of Eq.~(\ref{ros}) is equal to $+ 1$, if there is a turning point of light, otherwise it is equal to $- 1$. 
\begin{figure}
  \includegraphics[width=0.7\linewidth]{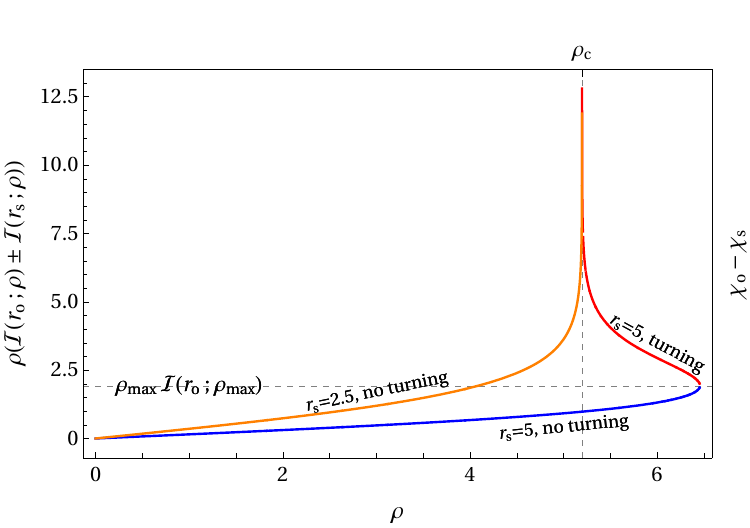}
  \caption{The $\chi_{\text{o}} - \chi_{\text{s}}$ as function of $\rho$ for illustrating Eq.~(\ref{ros}). It shows that the $\rho$ can be uniquely determined with given $\chi_{\text{o}} - \chi_{\text{s}}$ and $r_\text{s}$. Here, we set $\Lambda=10^{-4}M^{-2}$.\label{F4}}
\end{figure} 

For illustration, we show $\rho (\mathcal{I} (r_\text{o} ; \rho) \pm \mathcal{I} (r_\text{s} ;
\rho))[=\chi_{\text{o}} - \chi_{\text{s}}]$ as function of $\rho$ in Figure~\ref{F4}. The plot for Schwarzschild black hole also can be found in Ref.~\cite{Kocherlakota:2024hyq}. For a source outside the
photon sphere, there is a turning point, if following 
condition satisfies
\begin{eqnarray}
  \chi_{\text{o}} - \chi_{\text{s}} & \geq & \rho_{\max} \mathcal{I} (r_\text{o}
  ; \rho_{\max}) \nonumber\\
  & = & \frac{2 \sqrt{2}}{r_{\text{s}}} \sqrt{\frac{r_{\text{s}} - 2 M}{6 M -
  r_{\text{s}} + \sqrt{\left( r_{\text{s}} - 2 M \right) \left( r_{\text{s}} +
  6 M \right)}}} F \left( \arcsin \sqrt{\frac{r_2^\ast (r -
  r_3^\ast)}{r_3^\ast (r - r_2^\ast)}}, \frac{r_3^\ast (r_2^\ast - r_1^\ast)}{r_2^\ast (r_3^\ast - r_1^\ast)} \right) ~. 
\end{eqnarray}
Rhs. of above equation is obtained based on $\mathcal{I} \left( r_{\text{s}} ;
\rho_{\max} \right) = 0$, which subsequently leads to
\begin{eqnarray}
  \rho_{\max} & = & \sqrt{\frac{3 r_{\text{s}}^3}{3 r_{\text{s}} -
  r_{\text{s}}^3 \Lambda - 6 M}} ~.
\end{eqnarray}
In the case of $\rho = \rho_{\max}$, the roots $r_{1, 2, 3}^\ast$ introduced in
Eq.~(\ref{Ir}) take the form of
\begin{eqnarray}
  r_3^\ast  = r_{\text{s}} ~, \hspace{0.5cm}
  r_2^\ast  =  - \frac{r_{\text{s}}}{2} + \frac{r_{\text{s}}}{2}
  \sqrt{\frac{r_{\text{s}} + 6 M}{r_{\text{s}} - 2 M}} ~, \hspace{0.5cm}
  r_1^\ast  =  - \frac{r_{\text{s}}}{2} - \frac{r_{\text{s}}}{2}
  \sqrt{\frac{r_{\text{s}} + 6 M}{r_{\text{s}} - 2 M}} ~.
\end{eqnarray}
It shows that the roots are independent of cosmological constant $\Lambda$. As shown in Figure~\ref{F4}, the value of $\rho$ and the presence of a turning point can both be uniquely determined by the given $\chi_\text{o} - \chi_\text{s}$. For higher-order images in the case of $\chi_\text{o}
- \chi_\text{s} > 2 \pi$,  sources outside the photon sphere have a turning point, whereas sources inside the photon sphere have no turning point.

We present the full procedure of determination of $(\varphi, \rho)$ with
given location of source and observers based on Eqs.~(\ref{phios}) and (\ref{ros}). The $(\varphi, \rho)$ can subsequently
determine the celestial coordinates defined in Eq.~(\ref{Image}). In conclusion, we can
establish a one-to-one mapping as following,
\begin{eqnarray}
  \text{RT} : \left( r_{\text{o}},  r_{\text{s}}, \theta_{\text{o}},
  \theta_{\text{s}}, \phi_{\text{s}} \right) & \mapsto & (\text{A-B type}, n, \varphi, \rho) ~,
\end{eqnarray}
where we have adopted the convention $\phi_\text{o} = 0$, and the range of $\phi_\text{s}$ is
extended to $(- \infty, \infty)$. For the static and spherical black holes, we show that the mapping RT can be the composition of two mappings, $\text{RT} =
\text{RT}^{(\rho)} \circ \text{RT}^{(\varphi)}$, namely
\begin{eqnarray}
  \text{RT}^{(\varphi)} &:& (\theta_\text{o}, \theta_\text{s}, \phi_\text{s})  \mapsto  (l, m, n,
  \varphi)~, \\
  \text{RT}^{(\rho)} &:& ( r_\text{o}, r_\text{s}, \theta_\text{o}, \theta_\text{s}, l, m, n, \varphi) 
  \mapsto  \rho ~. 
\end{eqnarray}
The one-to-one mapping also shows that, except for sources being swallowed by the black hole, all multiple images of an emission source in the space can theoretically reach the observer.

\begin{figure}
  \includegraphics[width=1\linewidth]{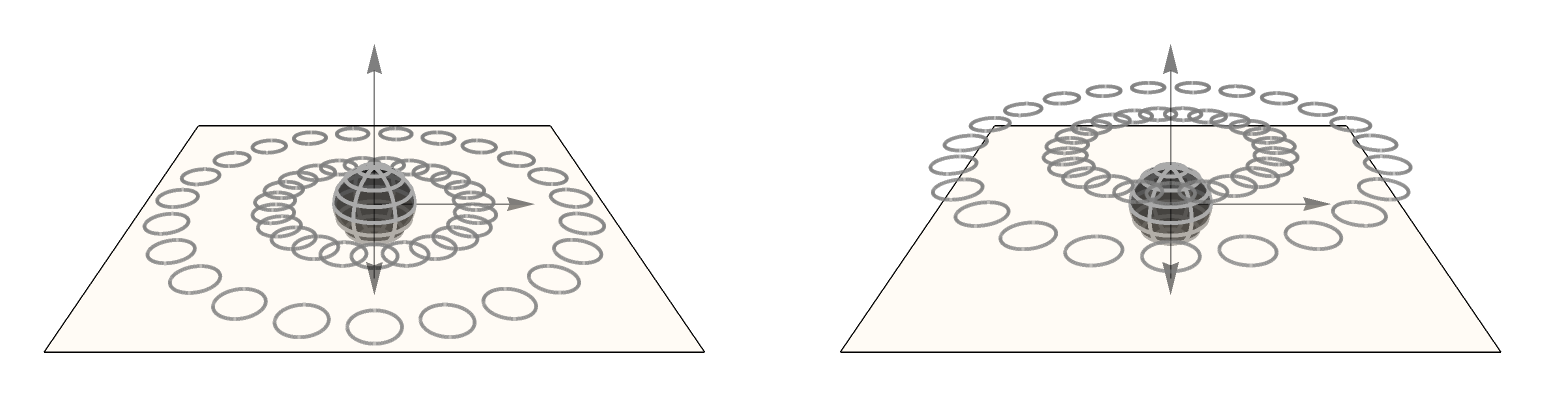}
  \caption{Discoid emission sources within (left panel) and above (right panel) the equatorial plane. \label{F5}}
\end{figure}

\begin{figure}
  \includegraphics[width=1\linewidth]{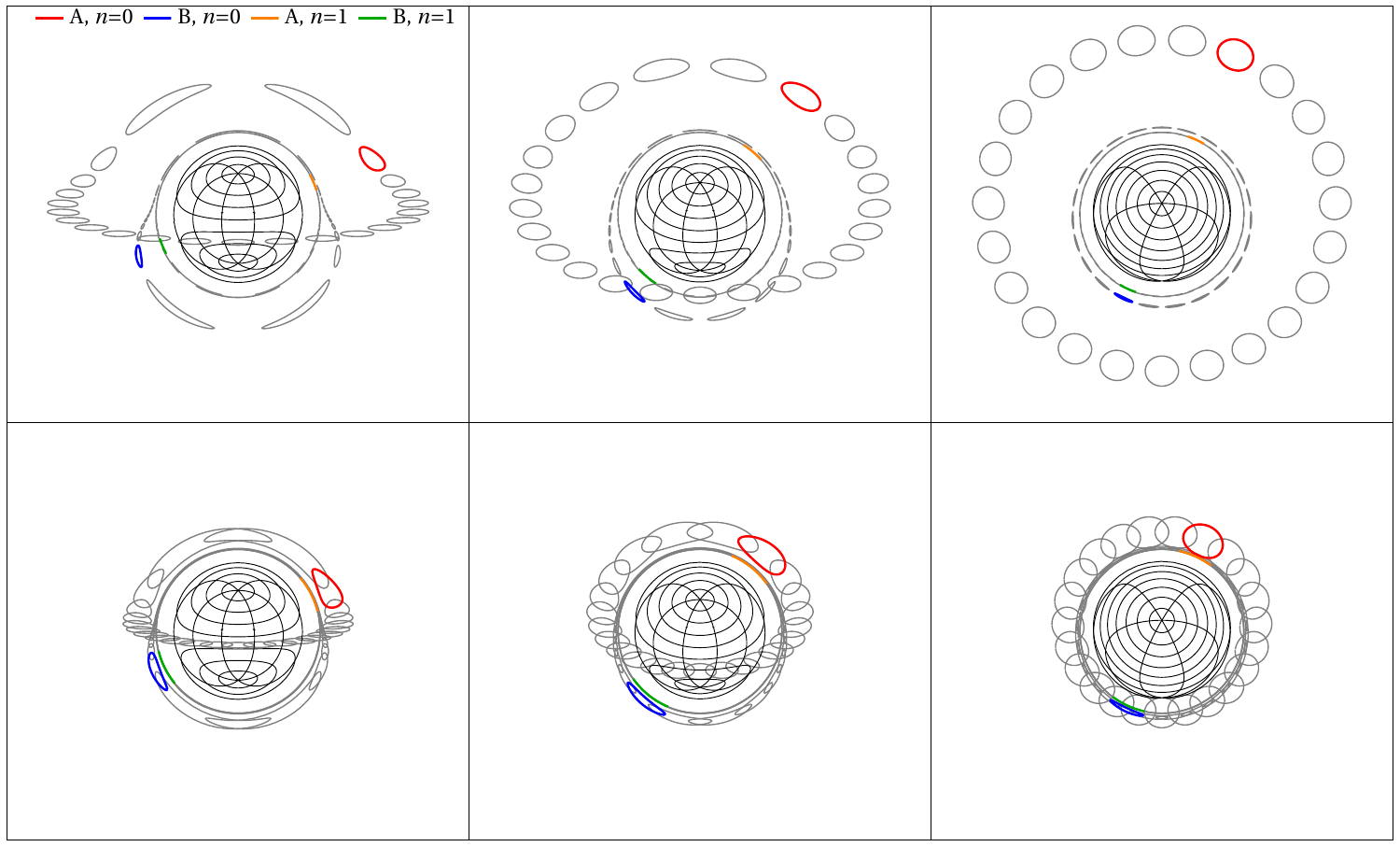}
  \caption{The multiple images of the emission sources given in the left panel of Figure~\ref{F5}. The observers are set to be the distance $r_\text{o}=500M$ and inclination angles are $\theta_\text{o}=80^\text{o}$ (left panel), $60^\text{o}$ (middle panel) and $20^\text{o}$ (right panel), respectively. Discoid sources located in the far and near regions are presented in the top and bottom panels, respectively. The cosmological constant is set to be $\Lambda=10^{-5}M^{-2}$.  \label{F6}}
\end{figure}

\begin{figure}
  \includegraphics[width=1\linewidth]{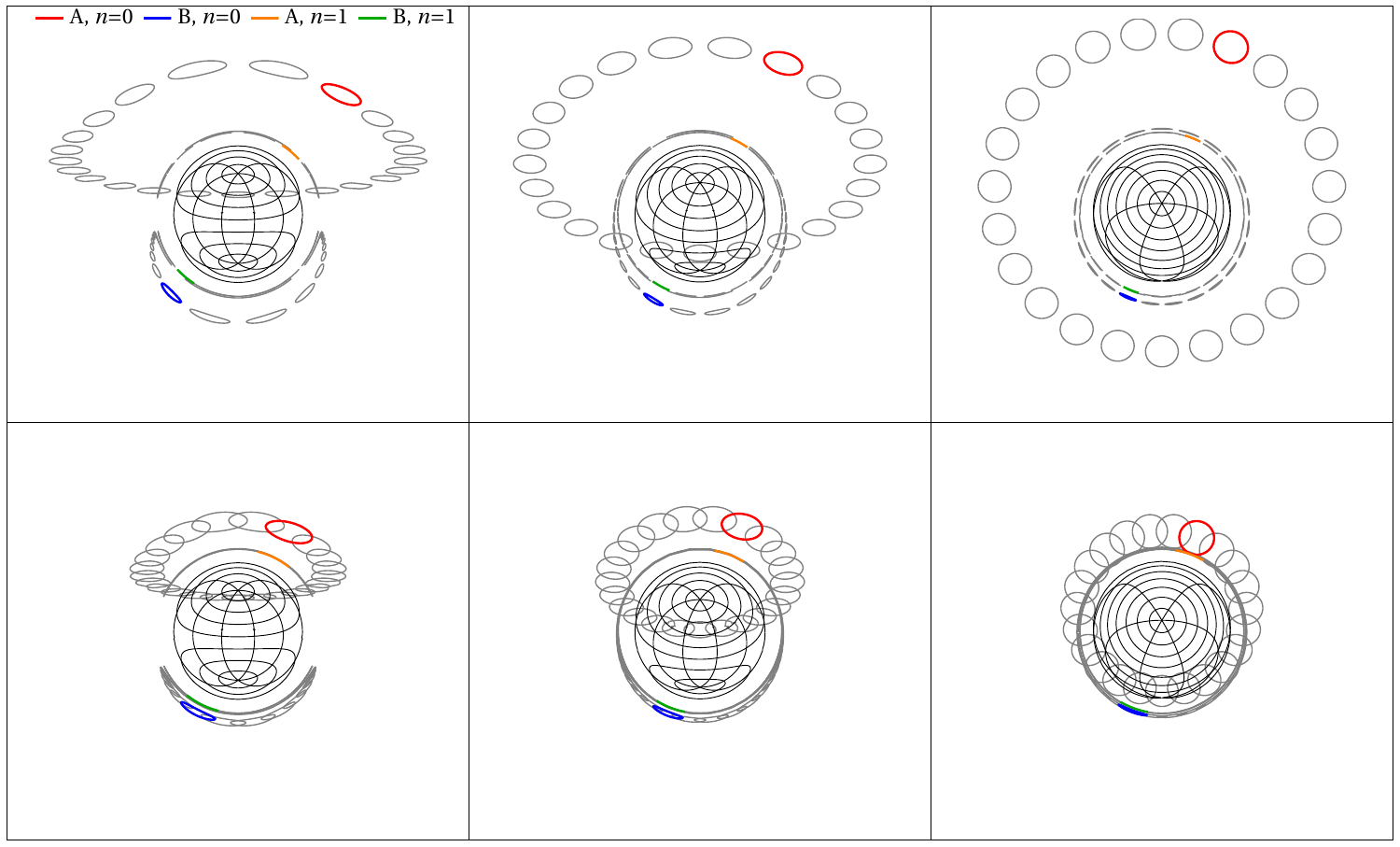}
  \caption{The multiple images of the emission sources given in the right panel of Figure~\ref{F5}. The observers are set to be the distance $r_\text{o}=500M$ and inclination angles are $\theta_\text{o}=80^\text{o}$ (left panel), $60^\text{o}$ (middle panel) and $20^\text{o}$ (right panel), respectively. Discoid sources located in the far and near regions are presented in the top and bottom panels, respectively. The cosmological constant is set to be $\Lambda=10^{-5}M^{-2}$. \label{F7}}
\end{figure}

In order to show the validation and efficiency of our ray tracing scenario, we consider discoid
emission sources on or above the equatorial plane in Figure~\ref{F5}. Utilizing our ray tracing, the images of these emission sources are presented in Figures~\ref{F6} and \ref{F7}. The A-type and B-type images are located on opposite sides relative to the center of the black hole. The points distributed near the surface of the horizon can all be imaged onto observers' sky, attributed to the strong gravity of the black holes. In principle, we can achieve infinite-precision simulations for the images, since the emission sources are projected directly onto the image plane.

\ 

\section{Signatures of moving hotspots in a thin disk \label{IV}}

The hotspot is a phenomenological model describing emission region distributed on the surface of accretion disks \cite{1992A257594B}. Recently, GRAVITY collaboration reported the flare events near the supermassive black hole \cite{GRAVITY:2018sef,GRAVITY:2023avo}, indicating that hotspots can serve as probes for horizon-scale physics.
In this section, we consider the moving hotspot in a thin disk, and study the
observational signature: temporal fluxes and
corresponding flux centroids. Besides the hotspots corotating with the accretion disk, we also theoretically study those in escape and plunging orbits \cite{Dovciak:2003jym,10.1093/pasj/55.1.155,Hackmann:2008zza}.

\subsection{Flux and centroid of hotspots}

The specific flux of a hotspot at frequency $\nu$ on image plane can be given by \cite{1992A257594B,GRAVITY:2020lpa,Li:2014coa}
\begin{eqnarray}
  F_\nu & = & \sum^{\infty}_{n = 0} \sum_{\textrm{X}}^{\{ A, B \}}
  F_\nu^{(   \textrm{X}, n )} = \sum_{n = 0}^{\infty} \sum_{\textrm{X}}^{\{ A, B \}} \int
  I_{\text{obs}} \left( \varphi^{( \textrm{X}, n )}, \rho^{( \textrm{X}, n )} ; t_{\text{o}},
  \textbf{x}_{\text{o}} \right) \textrm{d} \Omega^{( \textrm{X}, n )} ~, \label{Fnu}
\end{eqnarray} 
where the A and B represent the types of images defined in Eq~(\ref{rule}), $n$ is winding number of light orbits, and the solid angle $\textrm{d}\Omega^{( \textrm{X}, n )}$ is defined with the celestial coordinates on the sky, namely,
\begin{eqnarray}
  \textrm{d} \Omega & \equiv & \sin \Psi \textrm{d} \Psi \textrm{d} \Phi = \left( \frac{f
  (r_\text{o})}{r_\text{o} \sqrt{r^2_\text{o} - \rho^2 f (r_\text{o})}} \right) \rho \textrm{d} \rho \textrm{d}
  \varphi ~, \label{dOmega}
\end{eqnarray}
where we have ignored the superscript $( \textrm{X}, n )$, and the
second equal is obtained by making use of Eq.~(\ref{Image}). The
observed specific intensity $I_{\text{obs}}$ can be attributed to the redshifted
intensity of the emission sources, namely, \cite{Misner:1973prb,Cardenas-Avendano:2022csp}
\begin{eqnarray}
  I_{\text{obs}} \left( \rho, \varphi ; t_{\text{o}}, \textbf{x}_{\text{o}}
  \right) & = & g \left( \varphi, \rho, \textbf{x}_{\text{s}},
  \textbf{x}_{\text{o}} \right)^3 I_{\text{emt}} \left( t_{\text{s}},
  \textbf{x}_{\text{s}} \right) ~, \label{Iobs}
\end{eqnarray}
where the redshift factor $g$ is given by $g \equiv \nu_\text{o}/\nu_\text{s}  =  {(\left. u_{\text{o}} \cdot k
  \right|_{\textbf{x}_{\text{o}}})}/{(\left. u_{\text{s}} \cdot k
  \right|_{\textbf{x}_{\text{s}}})}$. For simplicity, we consider that the emission intensity $I_\text{emt}$ is isotropic and uniformly distributed across all frequencies, and that the environment around the black hole is optically thin. To quantify flux of the hotspots, one can also utilize the total flux, namely, $F=\int F_\nu\textrm{d}\nu$. In this case, corresponding total intensity is given by $I_\text{obs}^\text{tot}\equiv\int I_\text{obs}\textrm{d}\nu_\text{o}$, and Eq.~(\ref{Iobs}) can be evaluated in the form of $I_\text{obs}^\text{tot}=g^4 I_\text{emt}^\text{tot}$ \cite{1973ApJ...183..237C,1994ApJ...435...55B}. Here, we adopt the specific intensity at frequency $\nu$ as shown in Eq.~(\ref{Fnu}), because the VLBI telescopes are expected to operate at specific frequencies \cite{Emami:2022ydq,Johnson:2024ttr,Lu:2023bbn}. 

We consider hotspots moving within a thin disk located on the equatorial plane, where $\theta_{\text{s}} \equiv \pi/ 2$. In this case, Eqs.~(\ref{chiso}) and (\ref{phios}) reduce to  
\begin{subequations}
  \begin{eqnarray}
    \chi_\text{s} & = & \left( \frac{1}{2} - (2 n + l) \right) \pi ~,  \label{chisplane}\\
    \phi_\text{o} - \phi_\text{s} & = & - \arctan \left( \frac{\tan \varphi}{\cos \theta_\text{o}}
    \right) - \pi \text{sign} (\cos \varphi) \left( 2 n + l - m + (- 1)^m \left(
    \left[ \frac{\theta_\text{o}}{\pi / 2} \right] - \frac{1}{2} \right) \right)~. 
  \end{eqnarray} \label{chiphiplane}
\end{subequations}
For given $\varphi$, $\theta_\text{o}$ and $\theta_\text{s}$, we have $l^{\left( \text{B}
\right)} - l^{\left( \text{A} \right)} = 1$ according to 
Eq.~(\ref{rule}). Hence, Eq.~(\ref{chisplane}) can be rewritten as $\chi_\text{s} = (1 / 2 - \text{integer}) \pi$. It reduces to the result in Ref.~\cite{Zhu:2023kei}, where the author used $\chi_\text{s}$ to distinguish different lensing bands. 
By making use of Eqs.~(\ref{ros}), (\ref{dOmega}), and (\ref{chiphiplane}), the solid angle $\textrm{d} \Omega$ can be derived from the emission sources distributed on the surface of accretion disks, $r_\text{s} \textrm{d} r_\text{s} \textrm{d} \phi_\text{s}$, namely,
\begin{eqnarray}
  \textrm{d}\Omega&=& \left( \frac{\sin \Psi \textrm{d} \Psi \textrm{d}
  \Phi}{\rho \textrm{d} \rho \textrm{d} \varphi} \right) \left( \frac{\rho \textrm{d} \rho
  \textrm{d} \varphi}{r_\text{s} \textrm{d} r_\text{s} \textrm{d} \phi_\text{s}} \right) r_\text{s} \textrm{d} r_\text{s} \textrm{d} \phi_\text{s} \nonumber\\
   &=& \left( \frac{\partial}{\partial \rho} (\rho I_r) \right)^{- 1}
    \left( \frac{f (r_\text{o}) (1 - \sin^2 \theta_\text{o} \cos^2 \varphi)}{r_\text{o} \cos
   \theta_\text{o} \sqrt{r^2_\text{o} - \rho^2 f (r_\text{o})}} \right) \left( \frac{\rho^2}{r_\text{s}^2
   \sqrt{r_\text{s}^2 - \rho^2 f (r_\text{s})}} \right) r_\text{s} \textrm{d} r_\text{s} \textrm{d} \phi_\text{s}~, \label{dOmega2}
\end{eqnarray}
where $I_r \equiv \mathcal{I} (r_\text{o} ; \rho) \pm \mathcal{I} (r_\text{s} ; \rho)$. For primary images of corotating hotspots in a thin disk, Eq.~(\ref{dOmega2}) can reduce to the results in Ref.~\cite{1992A257594B}.

Based on Eqs.~(\ref{Iobs}) and (\ref{dOmega2}), the flux $F_\nu^{( \textrm{X}, n )}$ for a hotspot in a thin disk can be given by
\begin{eqnarray}
  F_\nu^{( \textrm{X}, n )} & = &  g \left( \textbf{x}_{\text{s}}, \textbf{x}_{\text{o}} ; \varphi, \rho
  \right)^3 I_{\text{emt}} \left( t_{\text{s}}, \textbf{x}_{\text{s}} \right)
  \Delta \Sigma \left( \frac{\partial}{\partial \rho} (\rho I_r) \right)^{- 1} \nonumber\\ && \times 
  \left. \left( \frac{f (r_\text{o}) (1 - \sin^2 \theta_\text{o} \cos^2 \varphi)}{r_\text{o} \cos
  \theta_\text{o} \sqrt{r^2_\text{o} - \rho^2 f (r_\text{o})}} \right) \left( \frac{\rho^2}{r_\text{s}^2
  \sqrt{r_\text{s}^2 - \rho^2 f (r_\text{s})}} \right) \right|_{\varphi = \varphi^{( \textrm{X}, n )} \left(\textbf{x}_{\text{o}},
  \textbf{x}_{\text{s}} \right), \rho = \rho^{( \textrm{X}, n )} \left(\textbf{x}_{\text{o}}, \textbf{x}_{\text{s}}  \right)} \label{flux}~,
\end{eqnarray}
where $\Delta \Sigma [\equiv r_\text{s} \textrm{d} r_\text{s} \textrm{d} \phi_\text{s}]$ describes the small area of a hotspot on the surface of the thin disk. The $\varphi^{( \textrm{X}, n )} $ and $\rho^{( \textrm{X}, n )}$ is obtained by making use of ray tracing scenario presented in Sec.~\ref{III}.

The physical time experienced by the observers is the proper time given by
\begin{eqnarray}
   \tau & = & t_\text{o} \sqrt{f (r_\text{o})} = \sqrt{f (r_\text{o})} (t_\text{s} +\mathcal{I}_t
  (r_\text{o} ; \rho) \pm \mathcal{I}_t (r_\text{s} ; \rho)) ~.
\end{eqnarray}
The second equal sign is given by evaluating the integral in Eq.~(\ref{intgeo2}),
namely, 
$
  t_\text{o} - t_\text{s}  =  \mathcal{I}_t (r_\text{o} ; \rho) \pm \mathcal{I}_t (r_\text{s} ; \rho)
$,
where the $\pm$ represent the turning point determined in 
Eq.~(\ref{ros}), and 
\begin{subequations}
  \begin{eqnarray}
    \mathcal{I}_t(r,\rho)\Big|_{\rho\geq\rho_c} &\equiv& -\frac{2 (r_1-r_3)}{\sqrt{r_3 (r_1-r_2)}} \Bigg(\frac{r_1^3 F\left(\arcsin\left(\sqrt{\frac{(r_1-r_2)
    (r-r_3)}{(r-r_1) (r_3-r_2)}}\right)\Big|\frac{r_1 (r_3-r_2)}{(r_1-r_2) r_3}\right)}{(r_1-r_{h_1}) (r_1-r_{h_2}) (r_1-r_{h_0}) (r_1-r_3)} \nonumber \\ &&
       -\frac{r_{h_2}^3 \Pi \left(-\frac{(r_{h_2}-r_1)
      (r_2-r_3)}{(r_1-r_2) (r_{h_2}-r_3)},\arcsin\left(\sqrt{\frac{(r_1-r_2) (r-r_3)}{(r-r_1) (r_3-r_2)}}\right)\Big|\frac{r_1 (r_3-r_2)}{(r_1-r_2) r_3}\right)}{(r_{h_1}-r_{h_2})
      (r_{h_2}-r_{h_0}) (r_{h_2}-r_1) (r_{h_2}-r_3)}\nonumber \\ && -\frac{r_{h_0}^3 \Pi \left(-\frac{(r_{h_0}-r_1) (r_2-r_3)}{(r_1-r_2) (r_{h_0}-r_3)},\arcsin\left(\sqrt{\frac{(r_1-r_2) (r-r_3)}{(r-r_1)
      (r_3-r_2)}}\right)\Big|\frac{r_1 (r_3-r_2)}{(r_1-r_2) r_3}\right)}{(r_{h_1}-r_{h_0}) (r_{h_0}-r_{h_2}) (r_{h_0}-r_1) (r_{h_0}-r_3)}\nonumber \\ && 
      \frac{r_{h_1}^3 \Pi \left(-\frac{(r_{h_1}-r_1) (r_2-r_3)}{(r_1-r_2) (r_{h_1}-r_3)},\arcsin\left(\sqrt{\frac{(r_1-r_2) (r-r_3)}{(r-r_1)
      (r_3-r_2)}}\right)\Big|\frac{r_1 (r_3-r_2)}{(r_1-r_2) r_3}\right)}{(r_{h_1}-r_{h_2}) (r_{h_1}-r_{h_0}) (r_{h_1}-r_1) (r_{h_1}-r_3)}\Bigg)~,\\
    \mathcal{I}_t(r,\rho)\Big|_{\rho<\rho_c} &\equiv&\frac{2 }{r_2-r_1}  \sqrt{\frac{(r_1-r_2) (r_1-r_3)^2}{r_3}}\Bigg(\frac{r_1^3 F\left(\arcsin\left(\sqrt{\frac{(r_1-r_2) (r-r_3)}{(r-r_1) (r_3-r_2)}}\right)\Big|\frac{r_1 (r_3-r_2)}{(r_1-r_2) r_3}\right)}{(r_1-r_{h_1})
    (r_1-r_{h_2}) (r_1-r_{h_0}) (r_1-r_3)} \nonumber \\ && -\frac{r_{h_2}^3 \Pi
     \left(-\frac{(r_{h_2}-r_1) (r_2-r_3)}{(r_1-r_2) (r_{h_2}-r_3)},\arcsin\left(\sqrt{\frac{(r_1-r_2) (r-r_3)}{(r-r_1) (r_3-r_2)}}\right)\Big|\frac{r_1 (r_3-r_2)}{(r_1-r_2)
     r_3}\right)}{(r_{h_1}-r_{h_2}) (r_{h_2}-r_{h_0}) (r_{h_2}-r_1) (r_{h_2}-r_3)}\nonumber \\ && -\frac{r_{h_0}^3 \Pi \left(-\frac{(r_{h_0}-r_1) (r_2-r_3)}{(r_1-r_2) (r_{h_0}-r_3)},\arcsin\left(\sqrt{\frac{(r_1-r_2) (r-r_3)}{(r-r_1) (r_3-r_2)}}\right)\Big|\frac{r_1 (r_3-r_2)}{(r_1-r_2) r_3}\right)}{(r_{h_1}-r_{h_0}) (r_{h_0}-r_{h_2}) (r_{h_0}-r_1)
     (r_{h_0}-r_3)}\nonumber \\ &&+\frac{r_{h_1}^3 \Pi \left(-\frac{(r_{h_1}-r_1) (r_2-r_3)}{(r_1-r_2) (r_{h_1}-r_3)},\arcsin\left(\sqrt{\frac{(r_1-r_2)
     (r-r_3)}{(r-r_1) (r_3-r_2)}}\right)\Big|\frac{r_1 (r_3-r_2)}{(r_1-r_2) r_3}\right)}{(r_{h_1}-r_{h_2}) (r_{h_1}-r_{h_0}) (r_{h_1}-r_1) (r_{h_1}-r_3)} \Bigg)~.
  \end{eqnarray}
\end{subequations}
The $r_{h_1}$, $r_{h_2}$ and $r_{h_3}$ are the roots of $f(r)=0$, in which the real roots represent horizons of Schwarzschild de-Sitter black holes, and $\Pi(x,y|z)$ is incomplete elliptic integral.
In the asymptotically flat spacetime, $f (r_\text{o} \rightarrow \infty) = 1$
indicates that coordinate time $t$ is sufficient for formulating the time experienced by distant observers. While, in case of  non-asymptotic flatness, $f (r_\text{o})
\neq 1$, one should adopt the proper time for the sake of rigor. The cosmological constant is shown to be a small quantity in the universe \cite{SupernovaSearchTeam:1998fmf}. For the supermassive black
hole in our galaxy, we have $f (r_\text{o}) - f_{\text{sch}} (r_\text{o}) = \Lambda r^2_\text{o} /
3 \sim 10^{- 12}$. For the supermassive black hole in M87, we have $f (r_\text{o}) -
f_{\text{sch}} (r_\text{o}) \sim 10^{- 5}$. The effect from cosmological constant
might not be ignored, if one expects to observe farther supermassive black holes
\cite{Johnson:2024ttr}. In this context, we will theoretically investigate whether the cosmological constant can affect the observed orbital periods of hotspots, as the observed time could be dilated or contracted.

Based on Eq.~(\ref{flux}), the flux centroid can be derived from the flux-weighted average of the positions, namely,
\begin{eqnarray}
  \textbf{C} & = & \mathcal{N}^{- 1}  \sum^{\infty}_{n =
  0} \sum_{\textrm{X}}^{\{ A, B \}} \left(\frac{F_\nu^{\left( n, \text{X} \right)}}{F_\nu}\right)
  \textbf{R}^{\left( n, \text{X} \right)} \left( \Psi^{\left( n, \text{X}
  \right)}, \Phi^{\left( n, \text{X} \right)} \right) \label{centro}
\end{eqnarray}
where $\textbf{R}$ is the 3-dimensional Cartesian coordinates of $(\Psi,
\Phi)$ in a unit sphere, and $\mathcal{N}$ is a normalization coefficient to
ensure $\textbf{C}$ to be a unit vector. In subsequent parts, we will transform the flux centroid $\textbf{C}$ onto the image plane using Eq.~(\ref{ImagePol}).

\

\subsection{Observational signatures of moving hotspots in different orbits}

\subsubsection{Circular motion}

Since the moving hotspot could be a phenomenological description of emission sources distributed on the surface of accretion disks, we consider it corotating around the black hole in circular orbits on the equatorial plane. The corresponding 4-velocities can take the form of
\begin{eqnarray}
  u^{(\text{cir})} & = & \sqrt{\frac{2}{2 f \left( r_{\text{s}} \right) -
  r_{\text{s}}   f' \left( r_{\text{s}} \right)}} \partial_t \pm
  \sqrt{\frac{f' \left( r_{\text{s}} \right)}{r_{\text{s}} \left( 2 f \left(
  r_{\text{s}} \right) - r_{\text{s}}   f' \left( r_{\text{s}} \right)
  \right)}} \partial_{\phi} ~.
\end{eqnarray}
For Schwarzschild black hole, the angular velocity of the circular orbit is
$w_{\text{K}} \equiv u^{(\text{cir}), \phi} / u^{(\text{cir}), t} = \left( m /
r^3_{\text{s}} \right)^{1 / 2}$. It is consistent with the results in
Newtonian gravity, and thus could be called Keplerian motion.

Due to magnetohydrodynamic effect of accretion matter \cite{Dexter:2020cuv,Aimar:2023kzj,Baganoff:2001kw,Genzel:2003as,Yuan:2003dc,GRAVITY:2021hxs}, the hotspot might not undergo geodesic motion \cite{GRAVITY:2018sef}. We can consider the 4-velocity of a hotspot deviation from the Keplerian motion in the form of
\begin{eqnarray}
  u^{(\text{dK})} & = & \frac{1}{\sqrt{f \left( r_{\text{s}} \right) -
  \gamma^2 r_{\text{s}}^2 w^2_{\text{K}}}} \partial_t \pm \frac{\gamma
  w_{\text{K}}}{\sqrt{f \left( r_{\text{s}} \right) - \gamma^2 r_{\text{s}}^2
  w^2_{\text{K}}}} \partial_{\phi} ~, \label{udK}
\end{eqnarray}
where the parameter $\gamma > 1$ denotes the super-Keplerian motion, and $\gamma < 1$ denotes the sub-Keplerian motion.

\begin{figure}
  \includegraphics[width=1\linewidth]{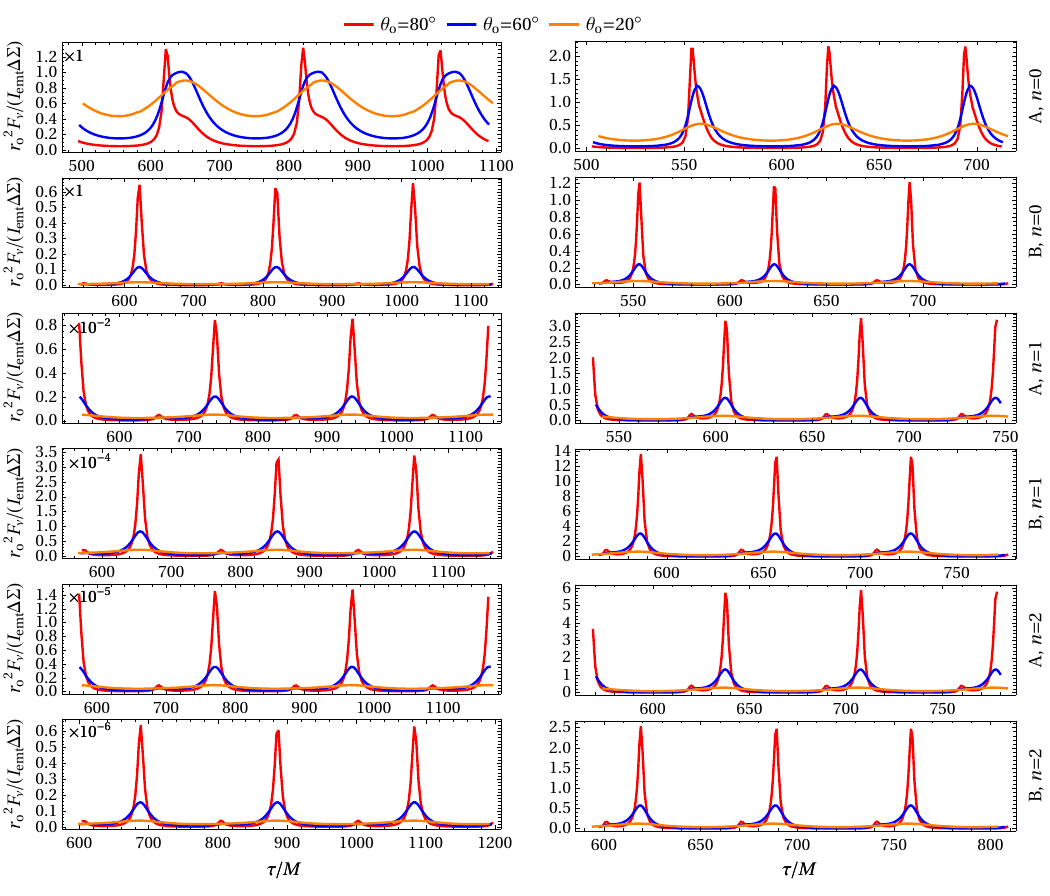}
  \caption{Temporal fluxes of corotating hotspots from primary to sixth-order images around Schwarzschild black hole for selected inclination angles. The sources move in circular orbits with fixed $r_\text{s}=10M$ (left panel) and $5M$ (right panel). The observers are set to be the distance $r_\text{o}=500M$.  \label{F8}}  
\end{figure}    
\begin{figure}
  \includegraphics[width=1\linewidth]{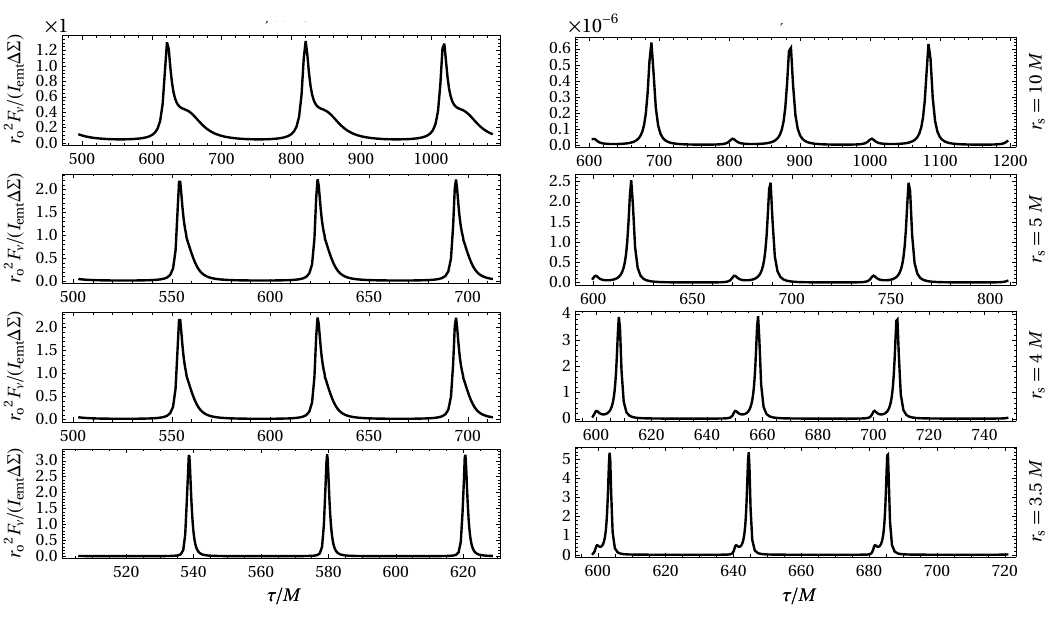}
  \caption{Temporal fluxes of corotating hotspots around Schwarzschild black hole for selected distance (denoted as $r_\text{s}$ in the plots) of the sources at inclination angle $\theta_\text{o}=4\pi/9$. We present primary images in left panel and sixth-order images in right panel. The observers are set to be the distance $r_\text{o}=500M$.  \label{F9}}  
\end{figure} 
\begin{figure} 
  \includegraphics[width=1\linewidth]{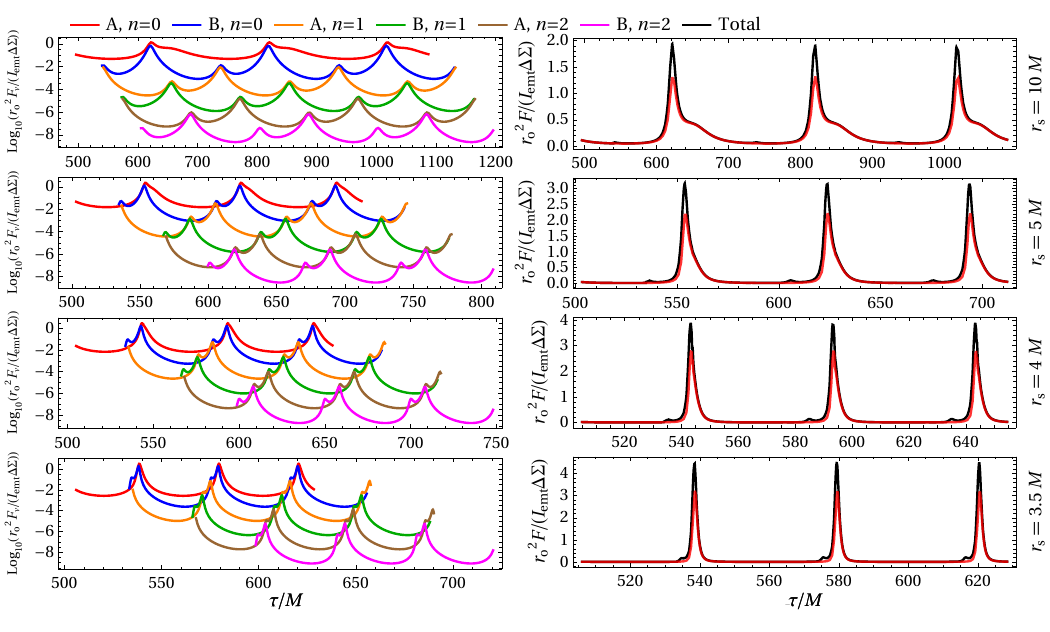}
  \caption{Temporal fluxes of corotating hotspots around Schwarzschild black hole for selected distance (denoted as $r_\text{s}$ in the plots) of the sources  at inclination angle $\theta_\text{o}=4\pi/9$. In the left panel, a log-plot is presented to illustrate the higher-order images, and in the right panel, the observed flux is compared with the flux from the primary image. The observers are set to be the distance $r_\text{o}=500M$.  \label{F10}}  
\end{figure}  
Figure~\ref{F8} shows temporal fluxes $F_\nu^{( \textrm{X}, n )}$ from primary to sixth-order images of the corotating hotspots. There is a time delay or advance between different order images. A large inclination angle sharpens peaks of the fluxes. 
The radii of hotspot orbit, denoted by $r_\text{s}$, also affect the profiles of temporal fluxes. Specifically, the magnitudes of the peaks are suppressed as the $r_\text{s}$  increases.
We further show the influence of the $r_\text{s}$  in Figure~\ref{F9}. For the sixth-order images, it is found that there are double peaks in the temporal fluxes. These two peaks get closer in time as the $r_\text{s}$ decreases. We compare temporal fluxes from primary to sixth-order images in Figure~\ref{F10}. One might find that the temporal fluxes exhibit self-similar profiles for higher-order images. And the secondary images are comparable to the primary ones and thus should not be neglected.

\begin{figure}
  \includegraphics[width=1\linewidth]{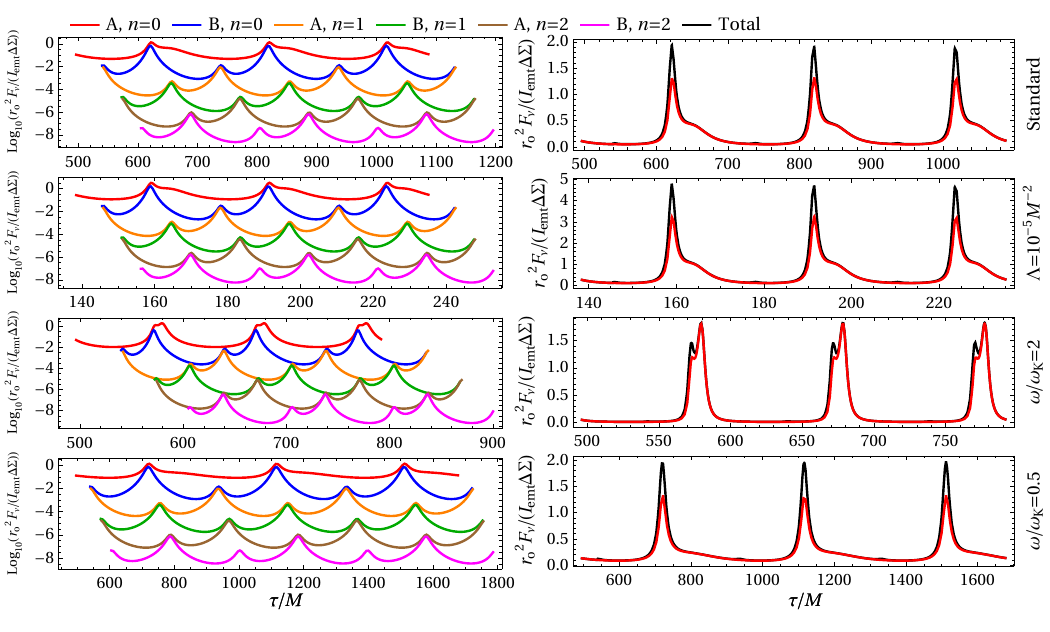}
  \caption{Temporal fluxes of corotating hotspots for the sources in the distance of $r_\text{s}=10M$ at inclination angle $\theta_\text{o}=4\pi/9$. We consider the influence from cosmological constant, super-Keplerian and sub-Keplerian orbits. In the left panel, a log-plot is presented to illustrate the higher-order images, and in the right panel, the observed flux is compared with the flux from the primary image. The observers are set to be the distance $r_\text{o}=500M$.  \label{F11}}  
\end{figure} 
\begin{figure}
  \includegraphics[width=1\linewidth]{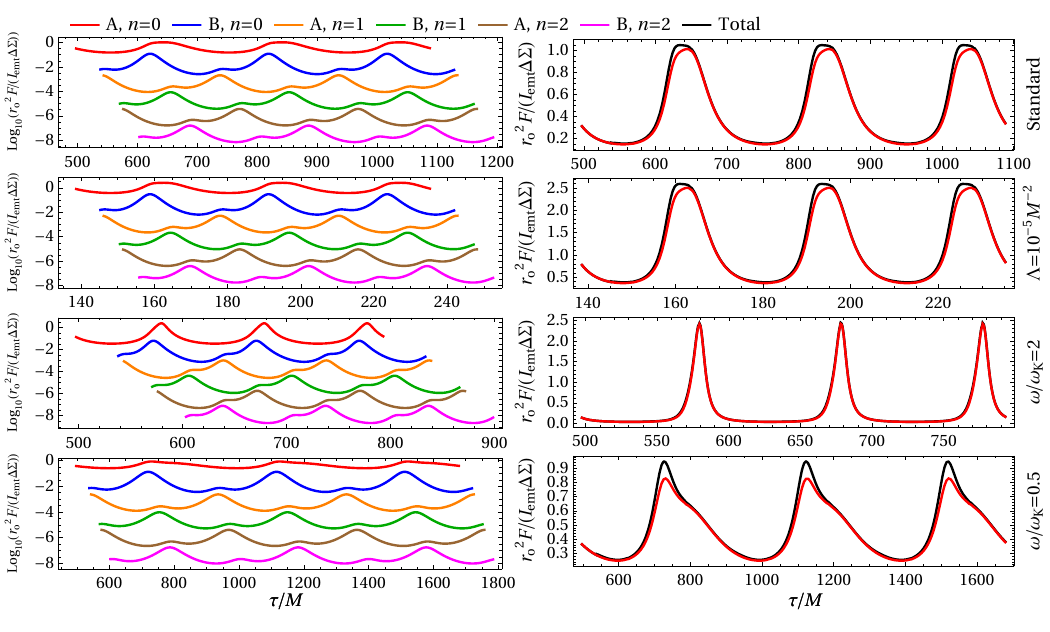}
  \caption{Temporal fluxes of corotating hotspots for the sources in the distance of $r_\text{s}=10M$ at inclination angle $\theta_\text{o}=\pi/3$. We consider the influence from cosmological constant, super-Keplerian and sub-Keplerian orbits. In the left panel, a log-plot is presented to illustrate the higher-order images, and in the right panel, the observed flux is compared with the flux from the primary image. The observers are set to be the distance $r_\text{o}=500M$.   \label{F12}}   
\end{figure}  
In Figure~\ref{F11} and \ref{F12}, we consider the temporal fluxes for super-Keplerian and sub-Keplerian motions, as
well as circular geodesic motion in the presence of cosmological constant. The $\gamma$ in Eq.~(\ref{udK}) and cosmological constant $\Lambda$ can both affect the period of the orbits.  In addition to the periods, it is found that super-Keplerian and sub-Keplerian orbits can significantly alter the profiles of temporal fluxes, particularly for the primary images. In contrast, the cosmological constant only affects the magnitudes but little affects profiles of the temporal fluxes. From the left panels of Figures~\ref{F10}, \ref{F11}, and \ref{F12}, one can observe that the classification of A-B type images is robust, as there are distinctive tendencies in the temporal fluxes, especially at the starting point.

\begin{figure}
  \includegraphics[width=1\linewidth]{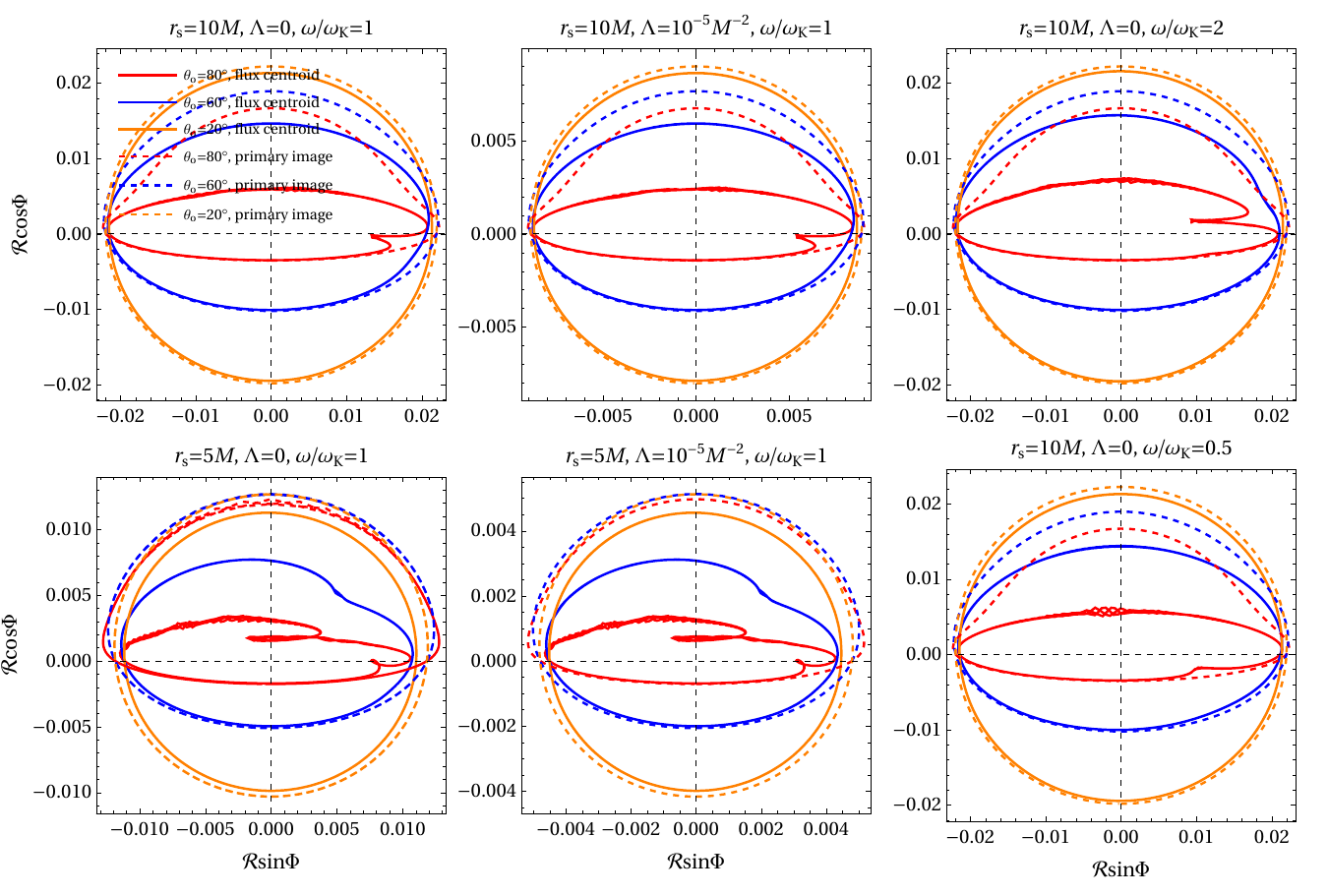}
  \caption{Flux centroids (solid) and tracks of primary images (dashed) of the corotating hotspots for selected inclination angles. The hotspot moves counterclockwise. We consider the influence from cosmological constant, super-Keplerian and sub-Keplerian orbits. The observers are set to be the distance $r_\text{o}=500M$.  \label{F13}}  
\end{figure} 
We present flux centroids of the corotating hotspots in super-Keplerian orbits, sub-Keplerian orbits, and circular geodesic orbits with and without a cosmological constant in Figure~\ref{F13}. The flux centroids are depicted as closed curves on the image plane. The super-Keplerian orbit leads to a more 
twisted centroid track, while the sub-Keplerian orbits can release the twist.
A non-vanishing cosmological constant alters size of the centroid track but does not affect its shape. More results about the influence of cosmological constant are presented in Appendix~\ref{appB}.
The centroid tracks are smaller than tracks of the primary images, because the higher-order images are closer to center of the black hole. The deviation between the centroid tracks and the tracks of the primary images becomes significant when the hotspots pass behind the black hole or when the hotspots are located near the black hole.

\

\subsubsection{Parameterized geodesic motion} 

\begin{figure} 
  \includegraphics[width=1\linewidth]{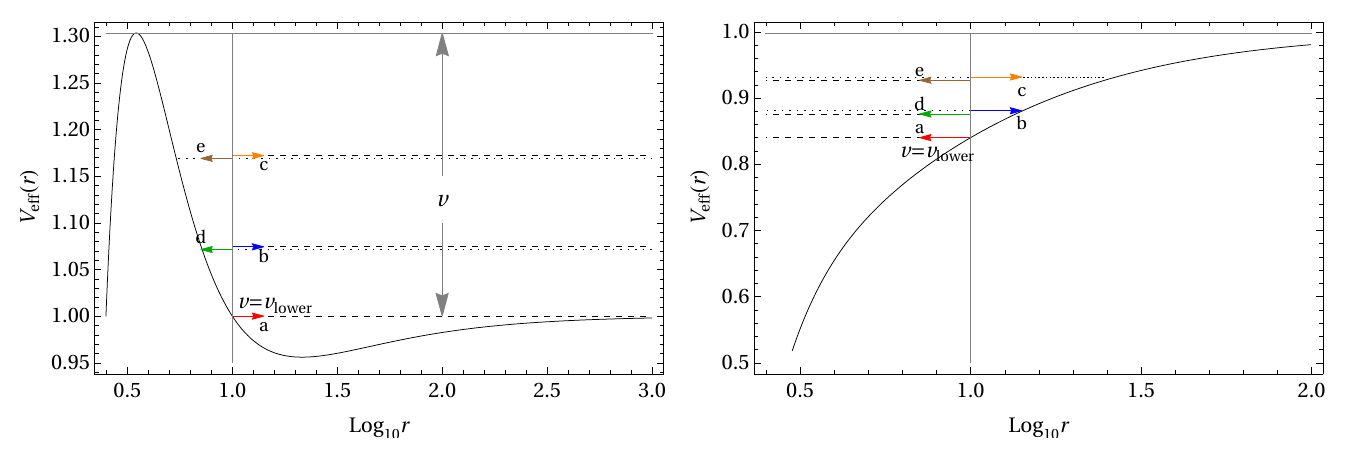} 
  \caption{Schematic diagrams of gravitational potential for illustrating the orbits that could have one turning points. We consider $L_0=5M$ (left panel) and $\sqrt{5}M$ (right panel).\label{F14}}  
\end{figure}

The hotspots might not be in a circular orbit, as the accretion flow can also be either swallowed or ejected by the supermassive black hole \cite{Dovciak:2003jym}. In order to study  signatures of moving hotspots with these features, we present a scenario for parameterizing all geodesic motions in a thin disk. Because the flare events reported by the GRAVITY collaboration all have a starting point on the image plane, we denote the initial location and initial speed of a flare event as $(r_0,\phi_0)$ and $v$. Thus, the 4-velocities of the hotspots can take the form of
\begin{eqnarray}
  u^{(\upsilon)} & = & \frac{1}{f (r)} \sqrt{\frac{f (r_0)}{1 - \upsilon^2}}
  \partial_r \pm \sqrt{\frac{f (r_0)}{1 - \upsilon^2} - f (r) \left( 1 +
  \frac{L_0^2}{r^2} \right)} \partial_r + \frac{L_0}{r^2} \partial_{\phi} ~, \label{uv}
\end{eqnarray}
where $L_0$ is the angular momentum, and the initial speed of a hotspot
can be derived from
\begin{eqnarray}
  \upsilon & = & \left. \sqrt{\frac{\gamma^{\ast} u \cdot
  \gamma^{\ast} u}{(u_{\text{stc}} \cdot u)^2}} \right|_{r = r_0} ~.
\end{eqnarray}
For geodesic motion, the angular momentum $L_0$ is conserved, while the
speed $v$ is not. 
Here, we use four parameters $(r_0, \phi_0, \upsilon, L_0)$  to quantify all types of geodesic orbits in the thin disk.
It is noted that $\upsilon \leqslant 1$, and there is a
lower bound of $\upsilon$ given by
\begin{eqnarray}
  \upsilon_{\text{lower}} & = & \frac{| L_0 |}{\sqrt{r_0^2 + L_0^2}} ~.
\end{eqnarray}
Motivated by the clockwise rotation of flare events around the galaxy center \cite{GRAVITY:2018sef,GRAVITY:2020lpa}, we consider orbits that could have one turning point, which results in an upper bound of $v$ formulated as
\begin{eqnarray}
  \upsilon_{\text{upper}} & = & \sqrt{1 - \frac{f (r_0)}{V_{\max}}} ~.
\end{eqnarray}
Figure~\ref{F14} illustrates the ranges of $v$ that we are interested. The escape and plunging orbits are presented in Figure~\ref{F15}, which will show to be sufficient to elucidate the observational signatures of the moving hotspots. In Appendix \ref{A}, we also provide additional orbits for examples.

\begin{figure}
  \includegraphics[width=1\linewidth]{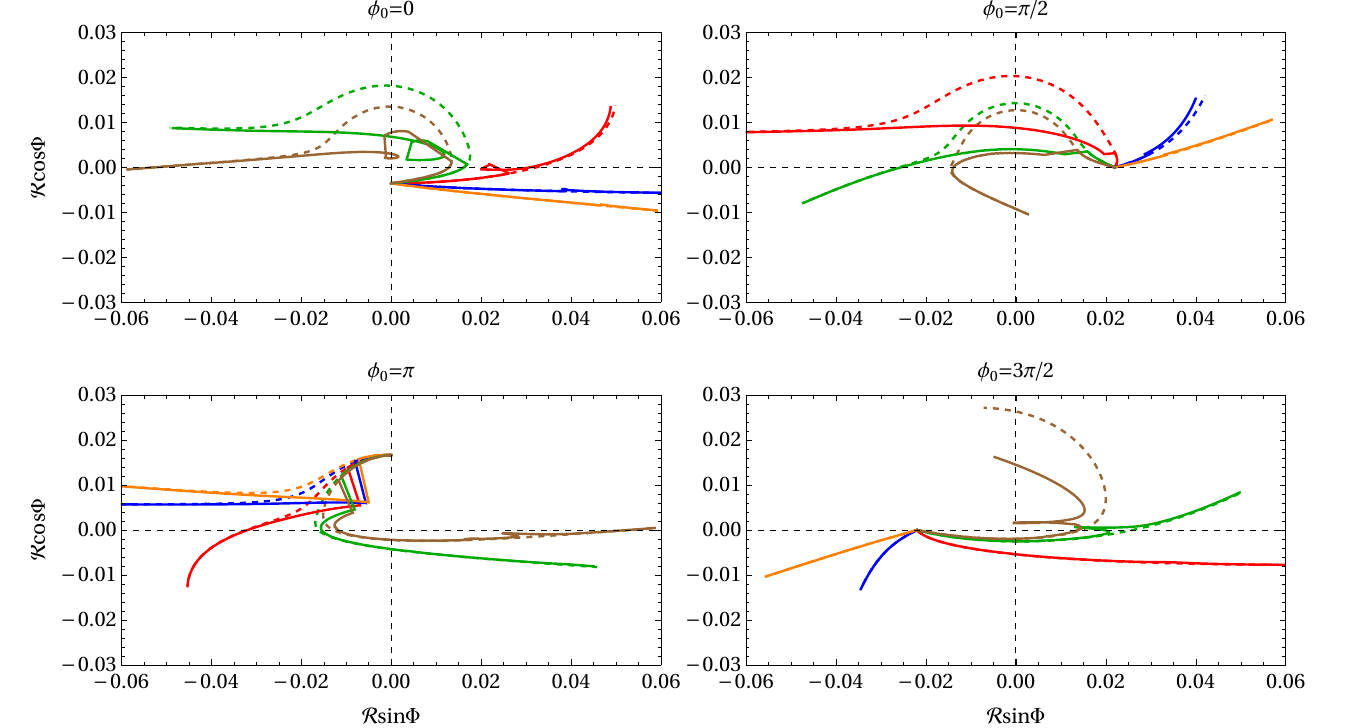}
  \caption{Flux centroids (solid curve) and tracks of primary images (dashed curve) of hotspots in escape orbits for selected $\phi_\text{o}$ and energy labeled in colored arrows in the left panel of Figure~\ref{F14}. The inclination angle is set to be $\theta_\text{o}=4\pi/9$.    \label{F16}}  
\end{figure}

\begin{figure}
  \includegraphics[width=1\linewidth]{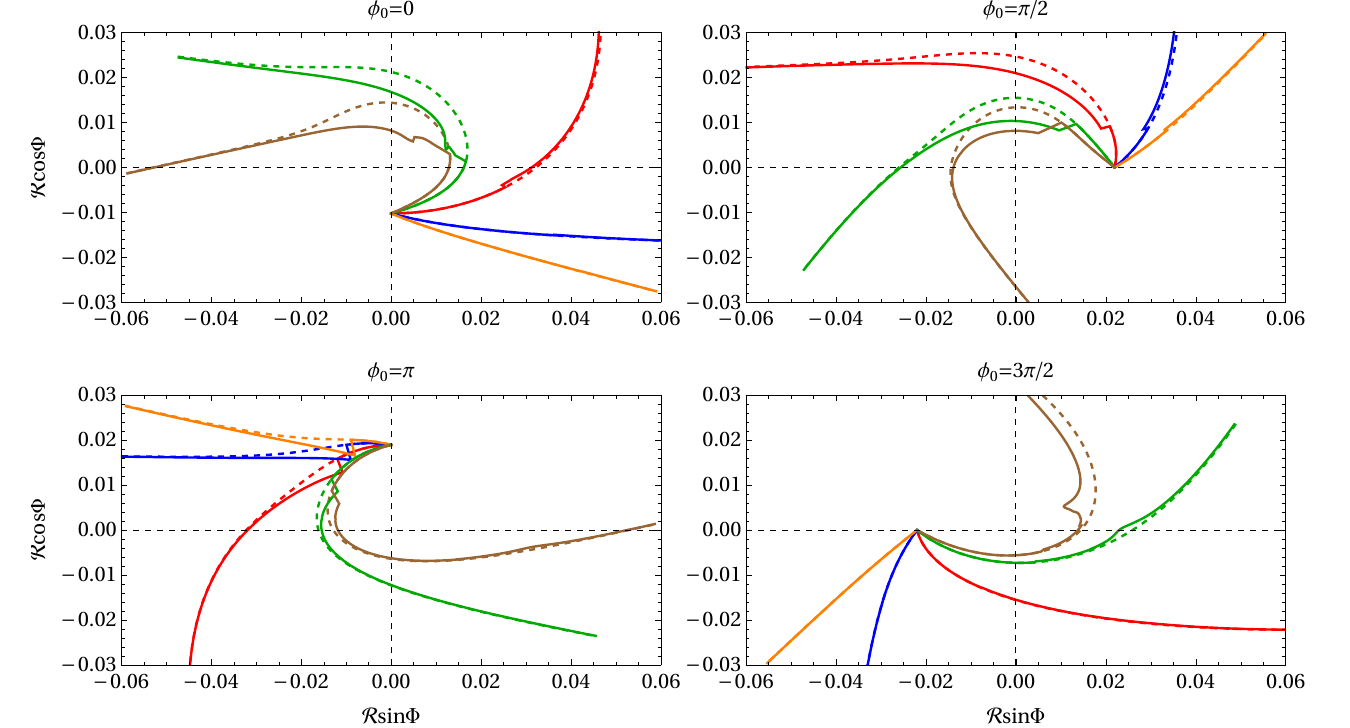}
  \caption{Flux centroids (solid curve) and tracks of primary images (dashed curve) of hotspots in escape orbits for selected $\phi_\text{o}$ and energy labeled in colored arrows in the left panel of  Figure~\ref{F14}. The inclination angle is set to be $\theta_\text{o}=\pi/3$ and the distance of observers is $r_\text{o}=500M$.    \label{F17}}  
\end{figure}

\begin{figure}
  \includegraphics[width=1\linewidth]{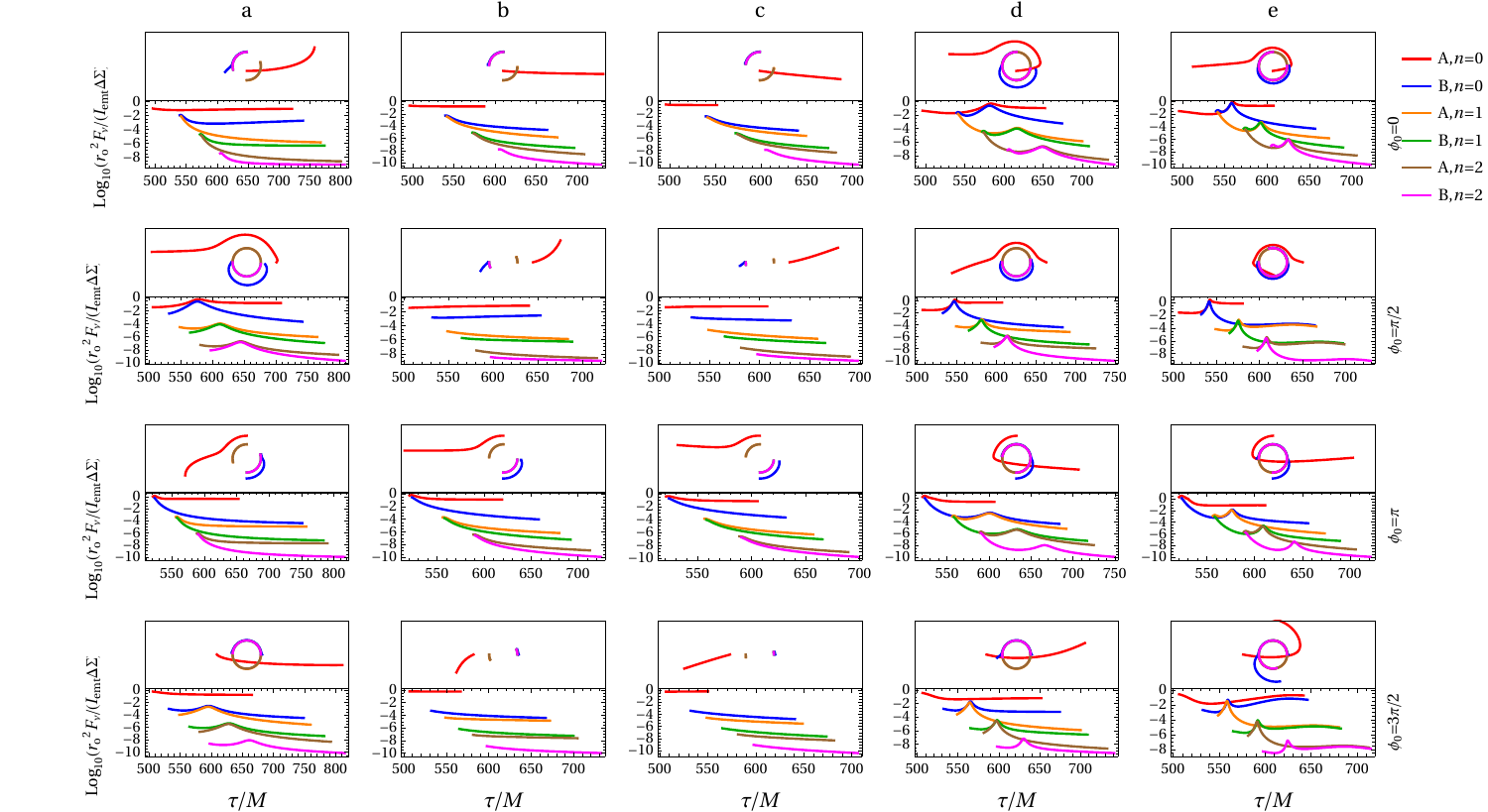}
  \caption{Tracks and temporal fluxes of primary to sixth-order images for the hotspots in escape orbits labeled in colored arrows in the left panel of Figure~\ref{F14}. The inclination angle is set to be $\theta_\text{o}=4\pi/9$ and the distance of observers is $r_\text{o}=500M$.   \label{F18}}  
\end{figure}

\begin{figure}
  \includegraphics[width=1\linewidth]{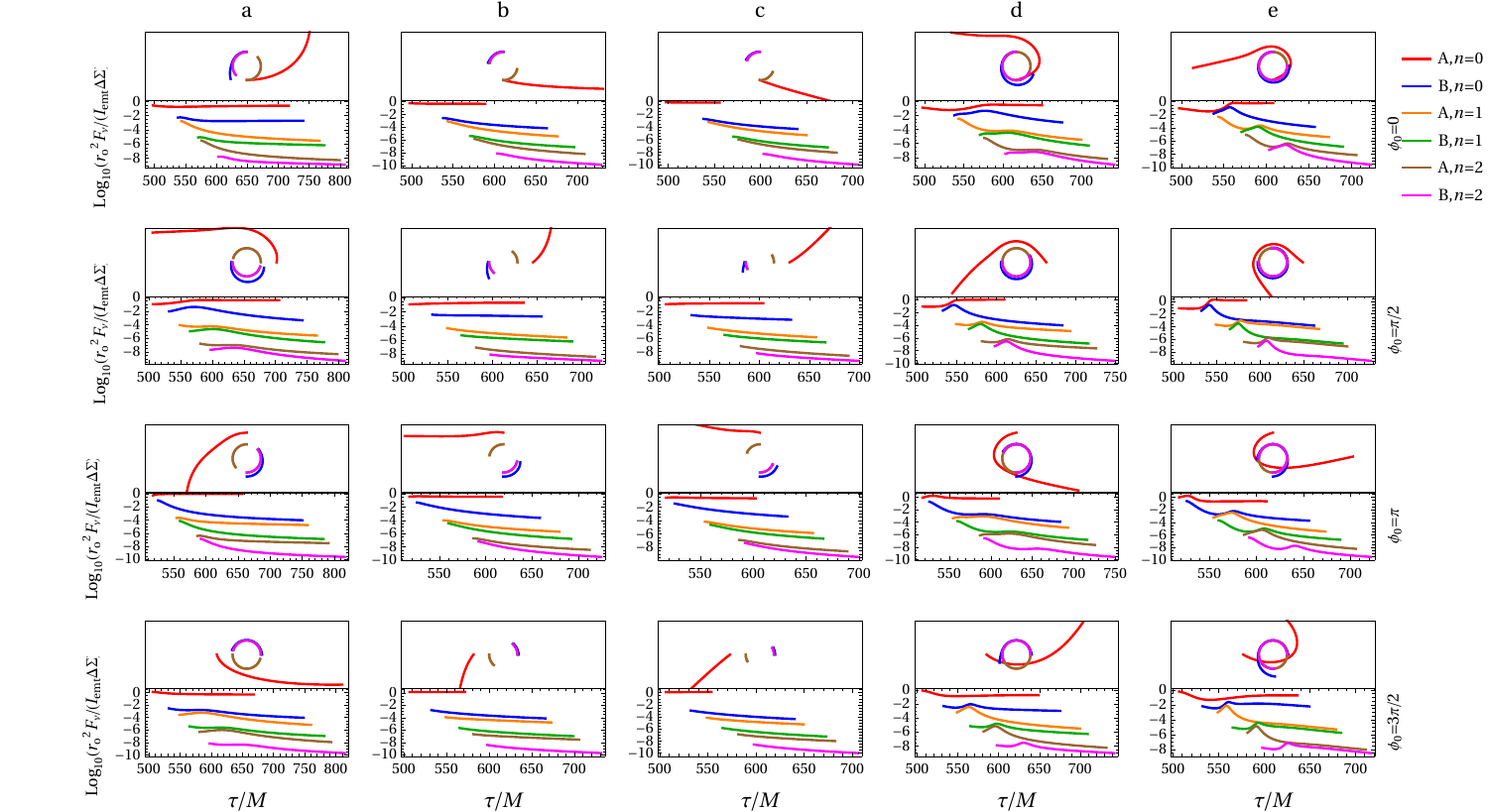}
  \caption{Tracks and temporal fluxes of primary to sixth-order images for the hotspots in escape orbits labeled in colored arrows in the left panel of Figure~\ref{F14}. The inclination angle is set to be $\theta_\text{o}=\pi/3$ and the distance of observers is $r_\text{o}=500M$.    \label{F19}}  
\end{figure}

Additionally, it is expected that the initial azimuth $\phi_0$ of the hotspots will influence the profiles of the temporal fluxes.
The centroid tracks of escape orbits for selected $\phi_\text{o}$ are shown in Figures~\ref{F16} and \ref{F17}. The sudden jumps in the tracks are caused by the emergence of secondary and higher-order images. Comparing the centroid tracks with corresponding tracks of the primary images, a significant difference is found when the hotspots pass behind the black hole. And a larger inclination angle $\theta_{\text{o}}$ appears to enhance the difference. In Figures~\ref{F18} and \ref{F19}, we present the tracks of the primary to sixth-order images and their corresponding temporal fluxes for the escape orbits given in left panel of Figure~\ref{F15}. As expected, the initial azimuth angle $\phi_\text{o}$ can affect temporal fluxes. A peak in the temporal flux is observed for orbits d and e, irrespective of the values of $\phi_\text{o}$. It could be attributed to the presence of a turning point in these orbits. 
\begin{figure}
  \includegraphics[width=0.8\linewidth]{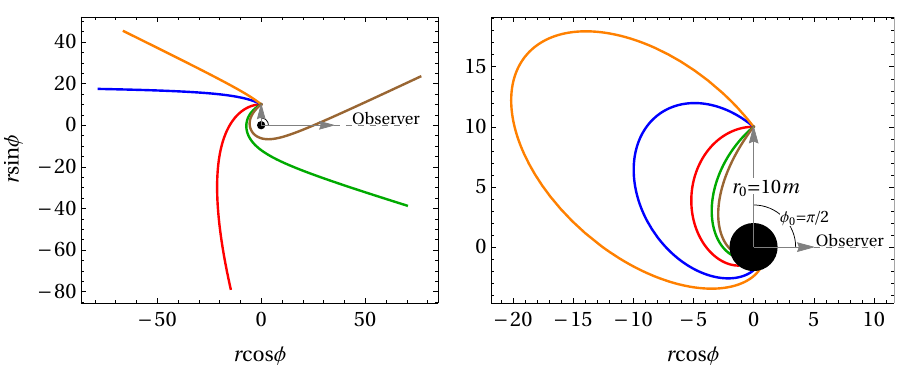}
  \caption{The trajectories of hotspots with given energy labeled in colored arrows in Figure~\ref{F14}.  \label{F15}}  
\end{figure} 

\begin{figure}
  \includegraphics[width=1\linewidth]{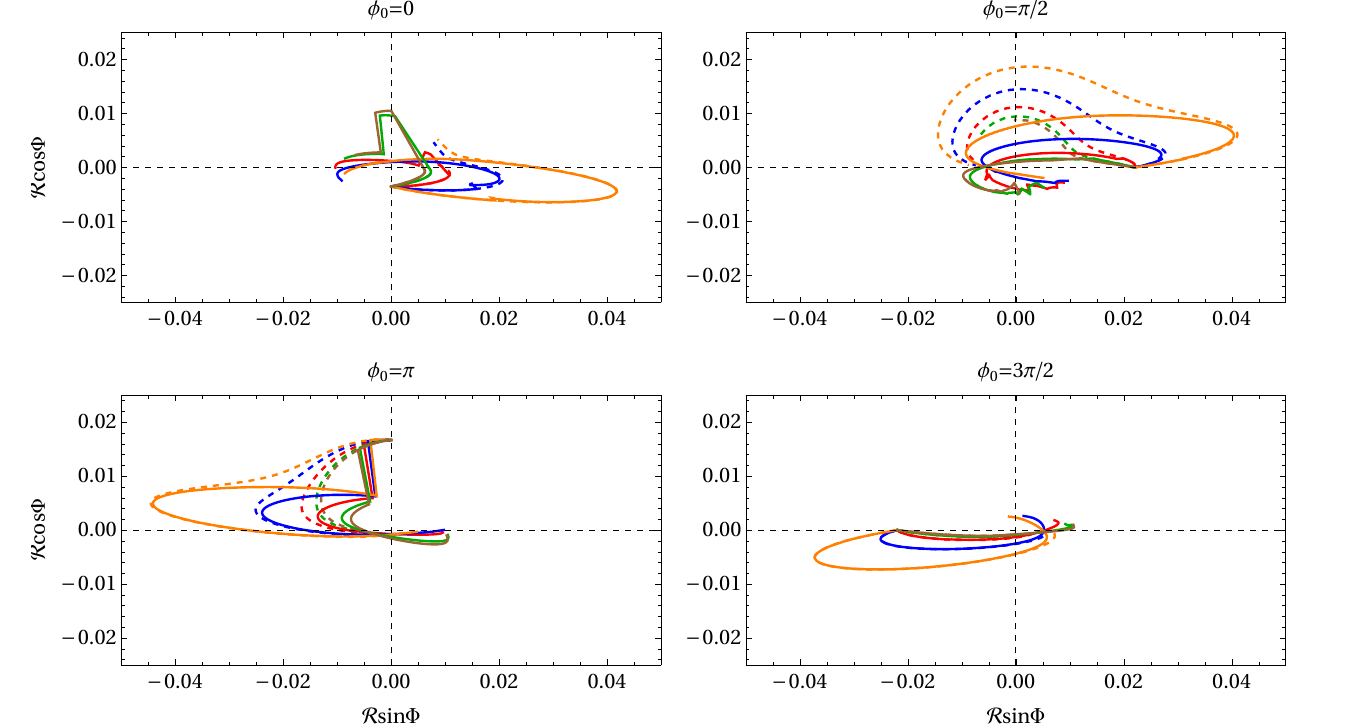}
  \caption{Flux centroids (solid curve) and tracks of primary images (dashed curve) of hotspots in plunging orbits for selected $\phi_\text{o}$ and energy labeled in colored arrows in the right panel of Figure~\ref{F14}. The inclination angle is set to be $\theta_\text{o}=4\pi/9$.    \label{F20}}  
\end{figure}

\begin{figure}
  \includegraphics[width=1\linewidth]{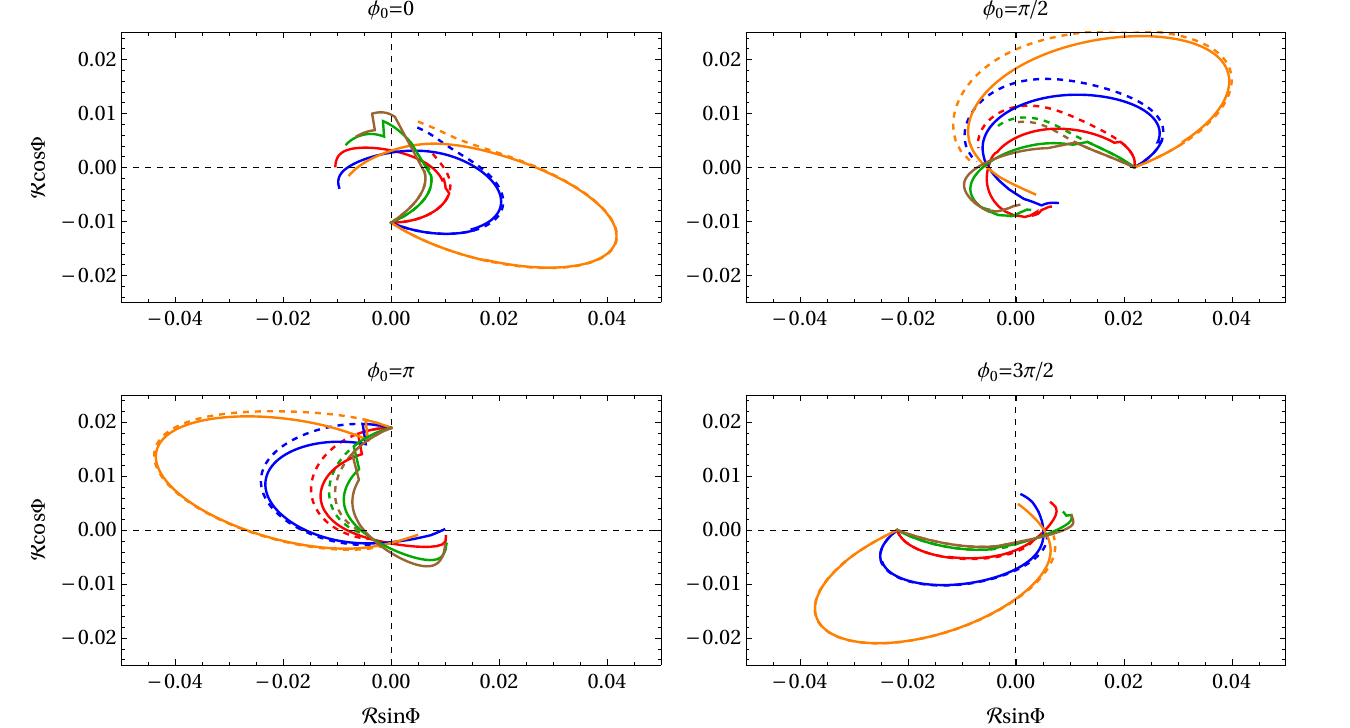}
  \caption{Flux centroids (solid curve) and tracks of primary images (dashed curve) of hotspots in plunging orbits for selected $\phi_\text{o}$ and energy labeled in colored arrows in the right panel of Figure~\ref{F14}. The inclination angle is set to be $\theta_\text{o}=\pi/3$.   \label{F21}}  
\end{figure}

\begin{figure}
  \includegraphics[width=1\linewidth]{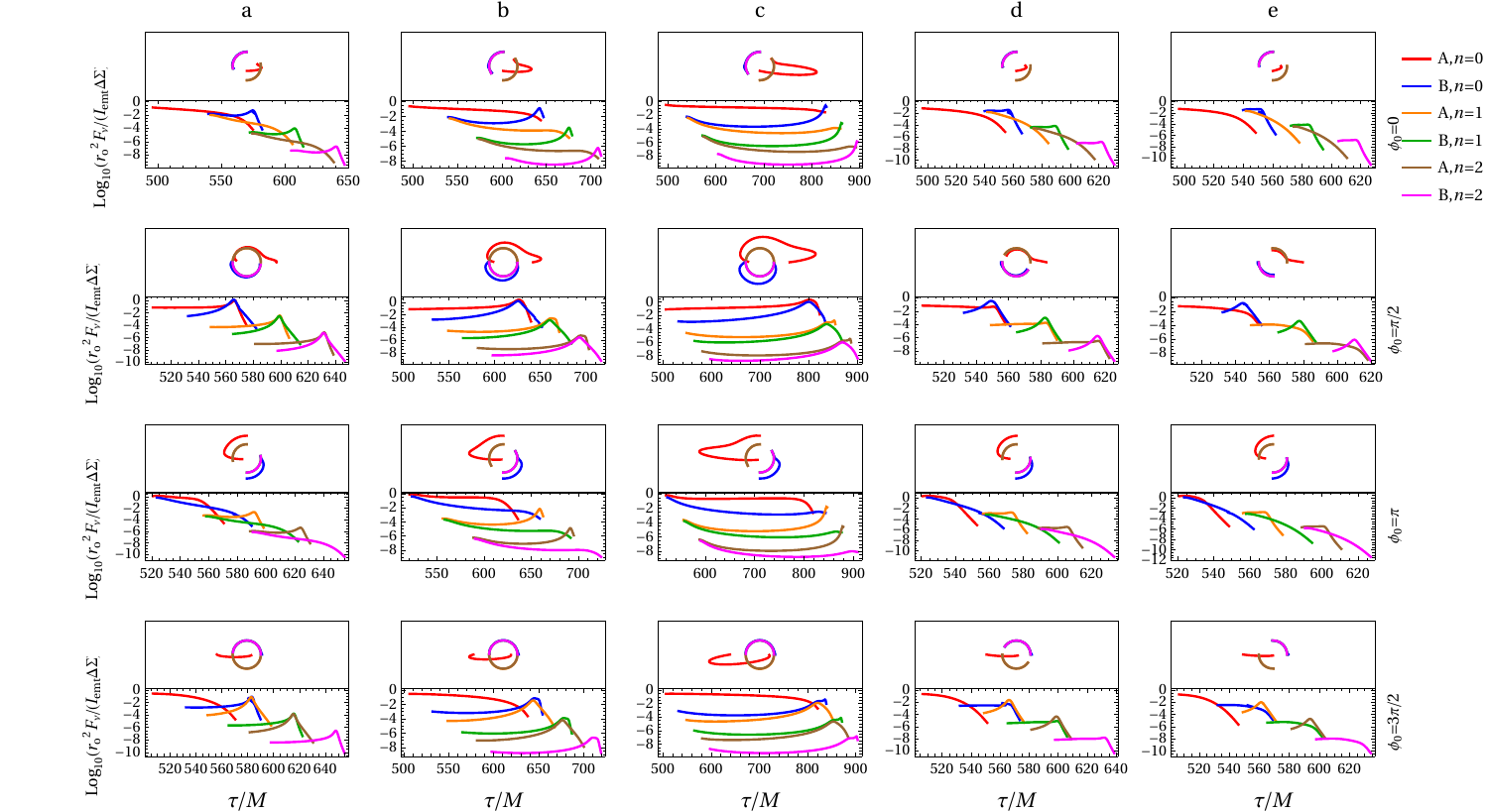} 
  \caption{Tracks and temporal fluxes of primary  to sixth-order images for the hotspots in plunging orbits labeled in colored arrows in the right panel of Figure~\ref{F14}. The inclination angle is set to be $\theta_\text{o}=4\pi/9$ and the distance of observers is $r_\text{o}=500M$.   \label{F22}}  
\end{figure}

\begin{figure}
  \includegraphics[width=1\linewidth]{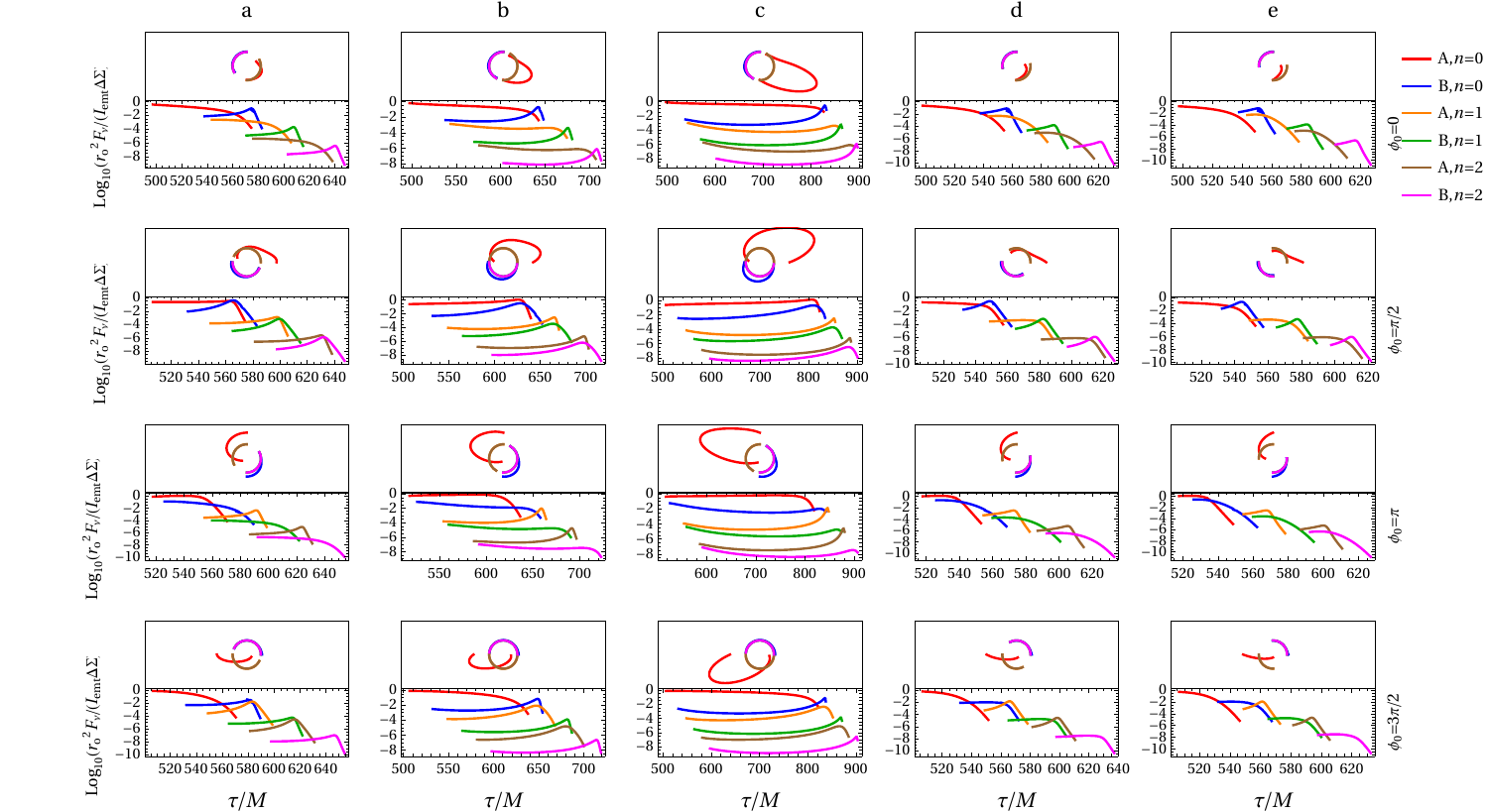}
  \caption{Tracks and temporal fluxes of primary  to sixth-order images for the hotspots in plunging orbits labeled in colored arrows in the right panel of Figure~\ref{F14}. The inclination angle is set to be $\theta_\text{o}=\pi/3$ and the distance of observers is $r_\text{o}=500M$.  \label{F23}}  
\end{figure}

In Figures~\ref{F20} and \ref{F21}, we present centroid tracks and tracks of the primary images, by considering hotspots in plunging orbits given in the right panel of Figure~\ref{F15}.  The tracks and temporal fluxes of primary to sixth-order images are presented in Figures~\ref{F22} and \ref{F23}. The swallowed hotspots seem to result in `echo' in the total temporal fluxes, as the flux decay into zero in an oscillatory manner. It occurs because 1) the fluxes from higher-order images can alternately dominate the observed flux after the hotspots are swallowed, and 2) one of the type A or type B images must exhibit a peak in the temporal flux.

As shown in Figures~\ref{F18}, \ref{F19}, \ref{F22} and \ref{F23}, one might find that the profiles of temporal fluxes are distinctive between images classified as A-type and B-type images. On equatorial plane, the A-type and B-type images correspond to even- and odd-order images, respectively. Their distinctive signatures were also found in the polarization images \cite{Himwich:2020msm}.

\section{Conclusions and discussions \label{V}}

This paper developed the ray tracing scenario as an extension of Ref.~\cite{1992A257594B}, to study higher-order images of moving hotspots near a spherical black hole. This scenario establishes a one-to-one mapping between the emission locations and the observer's sky for each image order.  We showed that a source located anywhere outside the black hole can be repeatedly mapped onto the observer's sky, from primary to higher-order images.  We presented the observational signatures from higher-order images of moving hotspots on the surface of a thin disk, around a Schwarzschild de-Sitter black hole. These hotspots are considered in circular, escape, and plunging orbits at various initial radii and initial azimuth angles.  We suggested that the higher-order images can be categorized into two types, denoted as A and B. Within each type, the temporal fluxes exhibit self-similar profiles. Temporal fluxes can be influenced by the angular velocities and orbital radii but are little affected by the cosmological constant. The flux centroid tracks are distinct from the tracks of the primary images, particularly when the hotspots pass behind the black hole.  

We adopted the astrometric approach to establish the observers' sky and image plane \cite{Chang:2020miq}. This approach is based on the principle that the observable should be defined with physical objectives, such as reference light rays in the universe. In previous studies~\cite{Chang:2020miq,He:2020dfo, Chang:2020lmg, Chang:2021ngy, Zhu:2023kei}, the reference light rays were set to those from the photon sphere. Here, we alternatively adopted a set of orthogonal light rays as references, which proves to be simple and concise.

In principle, our ray tracing scenario can achieve infinite-precision simulations for the images, because the emission sources are projected directly onto the image plane. Due to the strong gravity of the black holes, we showed that the points distributed near the surface of the horizon can all be imaged onto observers' sky. One can not see both the front and the back of an object simultaneously. However, this intuitive fact from a flat spacetime does not apply to a black hole. It also indicates that the hotspots can not be obscured by the central black hole. Nevertheless, due to the finite thickness of the accretion dicks, there is a self-eclipse region for the moving hotspots \cite{1992A&A...257..531K}, which consequently sharpens profile of the temporal flux. Thus, our study suggests that this eclipse effect might still exist for hotspots located anywhere outside black hole.

One might note that it is not rigorous to state that the hotspots are swallowed by the black hole, because the distant observer would see the hotspots frozen near the horizon.
As the hotspots approach the horizon shown in Figures~\ref{F22} and \ref{F23}, it was found that 1) the fluxes of higher-order images alternately dominate the observed flux, and 2) one of the type A or type B images must exhibit a peak in the temporal fluxes. These two points result in the temporal flux decaying with time in an oscillatory manner. This behavior is analogous to the amplitude decay of gravitational waves from an extremely compact object, known as gravitational wave echoes \cite{Cardoso:2017cqb,Mark:2017dnq}. 
 
Although one might suspect that the higher-order effect can not be detected by current observations \cite{EventHorizonTelescope:2019dse,EventHorizonTelescope:2022wkp}, studying the observational signatures of hotspots in higher-order images remains of theoretical interest \cite{Zamaninasab:2009df}, because it can reflect the intrinsic properties of black holes. Specifically, due to the secondary and tertiary images, the centroid track of a hotspot was shown to be differed from that of the primary images presented in Sec.~\ref{IV}. And the former, namely the simulated centroid position, was utilized in model fitting \cite{GRAVITY:2020lpa}. Since the centroid is mean position as shown in Eq.~(\ref{centro}), it inevitably depends on profiles of the temporal fluxes. In other words, the radiation mechanism can affect the centroids. A promising way to capture the effect from higher-order images is to average over multiple flares, as the hotspot model predicted a definite shift in the image centroids \cite{Broderick:2005at,Broderick:2005jj,Zamaninasab:2009df}. Additionally, the statistical methods might be an alternative way to extracted physical information about black holes \cite{1991A245454A,1994MNRAS.269..283Z,Berkley:2000mp,Pechacek:2008yf,Hadar:2020fda,Zhu:2023omf,Cardenas-Avendano:2024sgy}. With advancements in Very Long Baseline Interferometry (VLBI) in various wavelengths \cite{2012A&A...537A..52E,Lu:2023bbn}, and the proposals of new VLBI such as the ngEHT \cite{Emami:2022ydq} and BHEX \cite{Johnson:2024ttr}, the future observations are expected to achieve the resolution needed to address these challenge and shed light on the higher-order images near the supermassive black holes.
  
\ 

{\it Acknowledgments.} This work is supported by the National Natural Science Foundation of China under grants No.~12305073 and No.~12347101. The author thanks Prof. Xin Li and Mr. Jin-Tao Yao for useful discussions.

\

\appendix
\section{Signatures of moving hotspots in other orbits \label{A}}

\begin{figure}
  \includegraphics[width=1\linewidth]{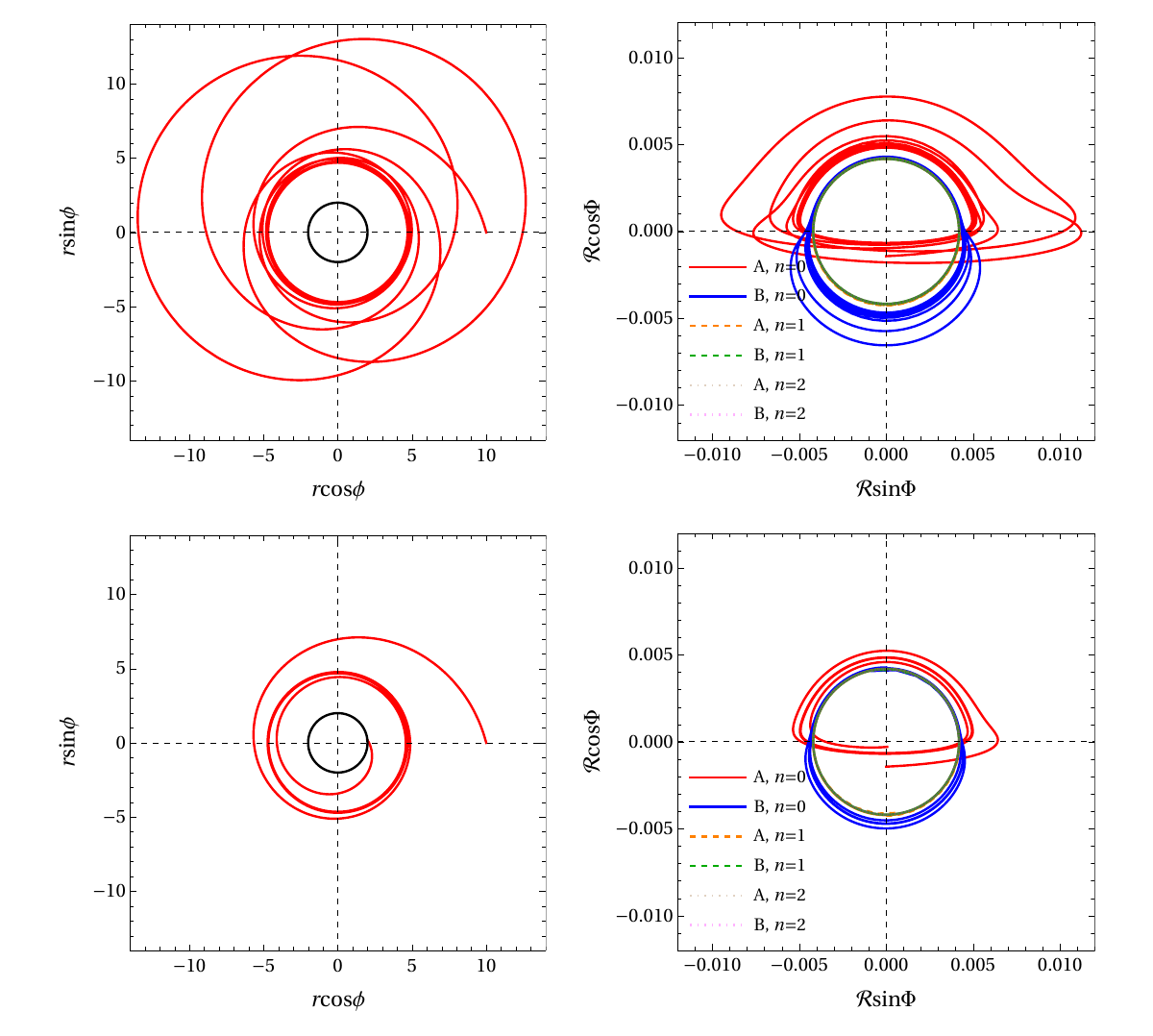}
  \caption{Left panel: trajectories of a hotspot in bounded orbit (top panel) and plunging orbit (bottom panel). Right panel: corresponding images of the hotspots.  The inclination angle is set to be $\theta_\text{o}=4\pi/9$, the distance of observers is $r_\text{o}=500M$, the cosmological constant is $\Lambda=10^{-5}M$.  \label{F24}}  
\end{figure}

\begin{figure}
  \includegraphics[width=0.9\linewidth]{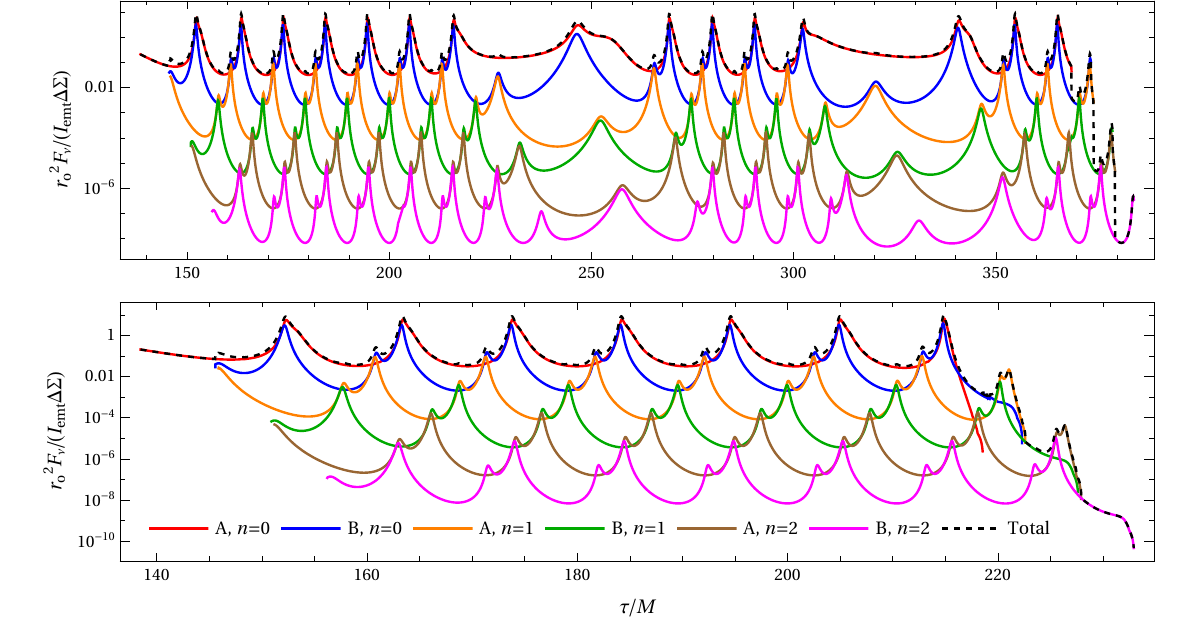}
  \caption{The temporal fluxes of hotspots in bounded orbit (top panel) and terminating orbit (bottom panel), as given in Figure~\ref{F24}.  \label{F25}}  
\end{figure}

\begin{figure}
  \includegraphics[width=0.85\linewidth]{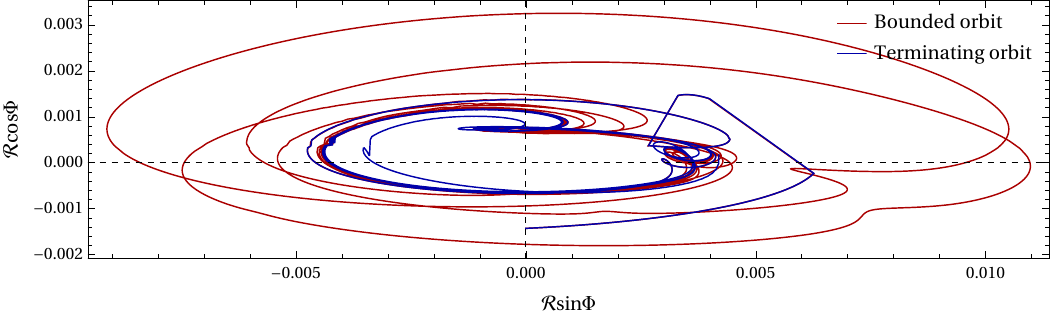} 
  \caption{The flux centroid of hotspots in bounded orbit and terminating orbit, as given in Figure~\ref{F24}. \label{F26}}  
\end{figure}

The escape, plunging and circular orbits studied in Section \ref{IV} appear to be simple, as they have no turning points, a single turning point, or a set of regular turning points. Therefore, we consider two representative orbits that have multiple turning points, based on 4-velocities parameterized in Eq.~(\ref{uv}). The trajectories and tracks from primary to sixth-order images are shown in Figure~\ref{F24}. The temporal fluxes and flux centroids are also presented in Figures~\ref{F25} and \ref{F26}, respectively. It shows that the observational signatures for the above two orbits are also found in the signatures for circular, escape, and plunging orbits, considered in Sec.~\ref{IV}. 


\section{Influence of cosmological constant \label{appB}}

Recent observations from the Dark Energy Spectroscopic Instrument (DESI) indicated a preference of the data for dynamical dark energy in our universe \cite{DESI:2024yrg}. It suggests that the cosmological constant might vary with scales, and there could be a distinct local cosmological constant in our galaxy. We thus considered a Schwarzschild-de Sitter black hole in our formalism in Sec.~\ref{III}. It might provide an interpretation of the super-Keplerian motion of the hotspots near Sgr A* as reported by GRAVITY collaboration \cite{Antonopoulou:2024qco,GRAVITY:2018sef,GRAVITY:2023avo}, because a positive cosmological constant might reduce the observed time, leading to shorter orbital periods of the hotspots.

\begin{figure}
  \includegraphics[width=0.9\linewidth]{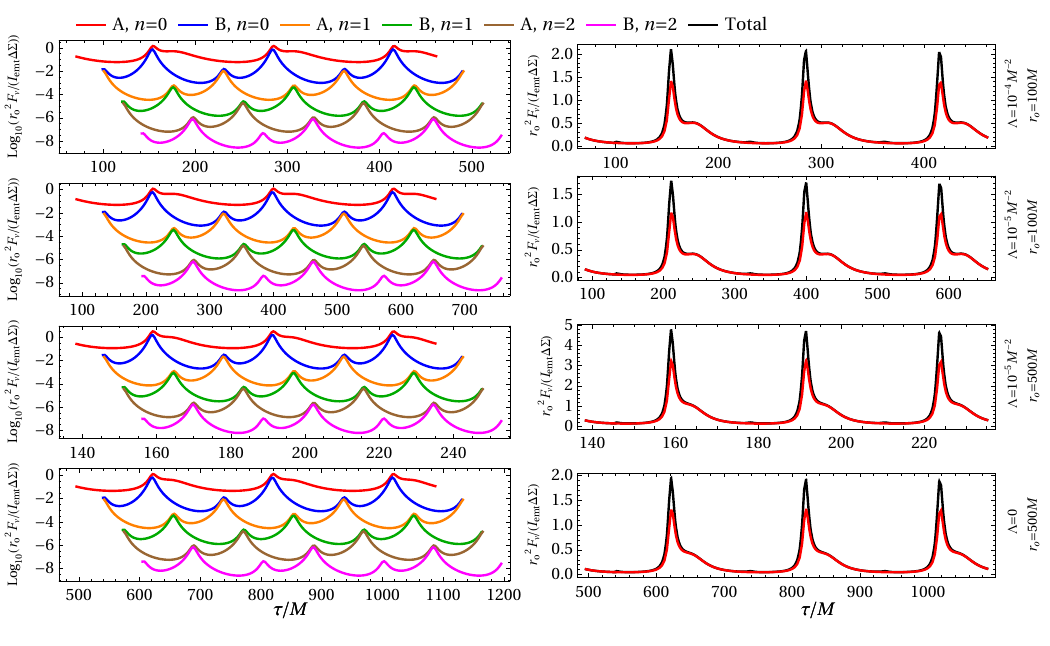}
  \caption{Temporal fluxes of corotating hotspots for selected cosmological constant and distance of observers.  In the left panel, a log-plot is presented to illustrate the higher-order images, and in the right panel, the observed flux is compared with the flux from the primary image. The distance of sources are set to be $r_\text{s}=10M$ and inclination angle is $4\pi/9$. \label{F27}}  
\end{figure}

\begin{figure}
  \includegraphics[width=0.85\linewidth]{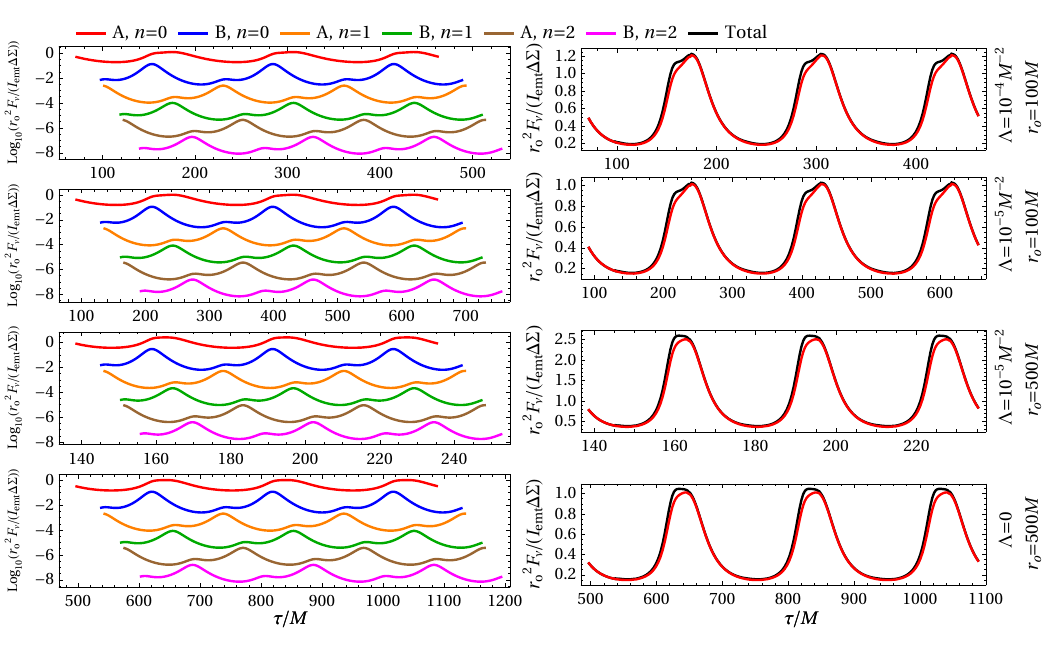} 
  \caption{Temporal fluxes of corotating hotspots for selected cosmological constant and distance of observers.  In the left panel, a log-plot is presented to illustrate the higher-order images, and in the right panel, the observed flux is compared with the flux from the primary image. The distance of sources are set to be $r_\text{s}=10M$ and inclination angle is $\pi/3$. \label{F28} }  
\end{figure}

In Figures~\ref{F27} and \ref{F28}, we present the temporal fluxes for selected values of cosmological constants. It shows that a larger positive cosmological constant leads to a shorter orbital period, as indicated by the closer spacing of flux peaks. Additionally, it is also found that the finite distance of observers can change the profile of temporal fluxes. The effect from the finite distance of observers might be addressed in the future studies.

\bibliography{cite}

\end{document}